\documentclass[11pt]{article}
\usepackage{amsfonts}
\usepackage{mathrsfs}
\usepackage{latexsym}
\usepackage{color}
\usepackage{amsmath,amssymb}
\usepackage{verbatim}
\definecolor{red}{rgb}{1,0,0}
\usepackage{cite}

\def\bea{\begin{eqnarray}} 
\def\eea{\end{eqnarray}}
\def\be{\begin{equation}} 
\def\ee{\end{equation}} 
\def\ba{\begin{array}}
\def\ea{\end{array}} 
\def\nn{\nonumber}

\begin{document}

\thispagestyle{empty}

\renewcommand{\thefootnote}{\fnsymbol{footnote}}
\setcounter{footnote}{0}
\setcounter{figure}{0}

\begin{center}
  
\vspace*{1.5cm}

{\Large\textbf{One-loop functional RG flow of scalar theory \\ with electroweak symmetry}\par} 

\vspace{1cm}

{\large Mahmoud Safari}

\vspace{5mm}

\textit{School of Particles and Accelerators,\\ Institute for Research in Fundamental Sciences (IPM),
P.O.Box 19395-5531, Tehran, Iran,\\ E-mail: msafari@ipm.ir}

\vspace{5mm}

\begin{abstract}
Using functional renormalization methods, we study the one-loop renormalization group evolution of theories with four scalars, at second order in the derivative expansion, in which electroweak symmetry is nonlinearly realized. In this framework we study the stability of $O(4)$ symmetry and find the $O(4)$-violating eigenperturbations and their corresponding eigenspectrum around three different geometries of the target space, namely those of the flat space, cylinder and sphere.
\end{abstract}

\end{center}

\vspace{1cm}

\section{Introduction}

Nonlinear realizations of symmetries play a central role in providing low-energy effective descriptions of theories with spontaneous symmetry breaking. Construction of these effective theories can be achieved in practice by the CCWZ prescription \cite{ccwz}. The chiral Lagrangian of strong interactions is an example of such an effective theory which describes pions arising as a result of the breakdown of chiral symmetry $SU(3)_L\!\times SU(3)_R$ of QCD to its vector subgroup by quark condensates. As another well known example, applying the CCWZ formalism to electroweak interactions leads to the so-called electroweak chiral perturbation theory. This gives, in terms of the Standard Model degrees of freedom, the most general effective theory based on electroweak symmetry breaking, $SU(2)_L\!\times\! U(1)_Y \rightarrow U(1)_{em}$. Its scalar sector consists of three Goldstone bosons, which are treated in the same way as pions, along with an extra scalar degree of freedom, the Higgs singlet.\\
Due to the presence of this `radial' Higgs mode, it is possible, at least locally, by making a field redefinition, to rewrite the theory in a way that the symmetry is actually realized linearly. The linearly transforming field is in fact given by the conventional parametrization of the Higgs field as an $SU(2)_L$ doublet $H$, in terms of which the scalar sector of the Standard Model effective Lagrangian with at most two derivatives reads
\be
\mathcal{L} = Z(\rho)\,\partial_\mu H^\dagger\partial^\mu H +V(\rho)+Y(\rho)\,\partial_\mu(H^\dagger H)\partial^\mu(H^\dagger H)+T(\rho)\,\vert H^\dagger\overset{{\hspace{0.5mm}\raisebox{-0.6mm}{\text{\scriptsize $\leftrightarrow$}}}}{\partial}_{\!\mu}\, H\vert^2\!\!, \hspace{5mm} \rho^2 = 2H^\dagger\! H, \label{lagrangian_H}
\ee
where the four functions $Z,V,Y$ and $T$ include terms with arbitrary powers of $H^\dagger\!H$. However, this is not always possible in a global manner. In this case we still say that the symmetry is nonlinearly realized \cite{percacci_safari}. This happens when the target space has a nontrivial topology. An example of this is provided by the minimal version of composite Higgs models with custodial symmetry \cite{mchm}, which is based on the $SO(5)/SO(4) = S^4$ symmetry breaking pattern, so that the target space has the topology of a four-sphere. 

In the absence of gauge fields and fermions, the electroweak chiral Lagrangian will lead to a theory of four scalars with electroweak symmetry, namely, $h$, a singlet of $SU(2)_L\!\times\! SU(2)_R$, and the unitary nonlinear sigma model field $U$
, parametrized by the three Goldstone bosons $\chi^\alpha$, which transforms as $U\rightarrow g_LUg_R^\dagger$ under $SU(2)_L\!\times\! SU(2)_R$. This effective Lagrangian, at second order in the derivative expansion, is 
\be \label{lagrangian_hU} 
\mathcal{L} = \frac{1}{2}\partial_\mu h \partial^\mu h +V(h)+ \frac{1}{4} K(h) \,\mathrm{Tr}\left(\partial_\mu U^\dagger\partial^\mu U\right)+\frac{1}{8}P(h)\left| \mathrm{Tr}\left(U^\dagger\partial_\mu U\sigma^3\right)\!\right|^2,
\ee
where $V(h)$ is the Higgs potential and the two functions $K(h)$ and $P(h)$ include all the couplings of the Higgs singlet to Goldstones. The couplings in $V(h)$ and $K(h)$ preserve $SU(2)_L\!\times\! SU(2)_R$ while those of the $P(h)$ function break it explicitly to $SU(2)_L\!\times\! U(1)_Y$. 

In this work we are interested in the renormalization of such theories in the nonlinear parametrization and at second order in the derivative expansion. The gauged Higgsless version has been studied in \cite{fptv}. Here instead, the presence of the singlet $h$ allows for an infinite number of couplings, collected into three independent functions $V(h)$, $K(h)$ and $P(h)$. Inclusion of the $O(4)$-violating couplings $P(h)$ also generalizes, at $N=4$, the $O(N)$ model studied in \cite{percacci_safari}. The quantity whose running we are interested in is the so-called effective average action (EAA). This is the same as the standard effective action which is the generator of $1$PI correlation functions, except that a cutoff term, bilinear in the fluctuating fields $\phi$,
\be  \label{cut_off}
\Delta S_k = \frac{1}{2}\int\!\! \frac{d^dp}{(2\pi)^d} \,\phi(-p) R_k(p^2) \phi(p),
\ee
is included in the path integral, which effectively cuts the integrated momenta at an IR scale $k$. 
The cutoff kernel $R_k$ is required to be a decreasing function of $p^2$ which vanishes in the large $p^2/k^2$ limit and tends to infinity for large values of the scale $k$. 
This is much in the spirit of Wilson's idea of renormalization \cite{wilson}. The cutoff term \eqref{cut_off} is finally subtracted from the effective action in order for it to reproduce the correct UV behaviour. The one-loop EAA is thus given by 
\be  \label{one-loop-eea}
\Gamma_k^{1\!-\!\mathrm{loop}} = S + \frac{1}{2}\,\mathrm{Tr}\log \left(\frac{\delta^2S}{\delta\phi\delta\phi}+R_k\right),
\ee
where $S$ is the tree level action. This is similar in structure to the familiar one-loop expression, except for the appearance of the cutoff kernel $R_k$ which was introduced to suppress the low momentum fluctuations. The properties of the cutoff kernel guarantee that the scale dependent effective action \eqref{one-loop-eea} interpolates between the tree level action in the UV and the one-loop effective action with complete integration of momenta in the IR. Deriving equation \eqref{one-loop-eea} with respect to $t=\log k$ gives the one-loop beta functional of the EAA
\be \label{1loop_flow}
\partial_t\Gamma^{1\!-\!\mathrm{loop}}_k = \frac{1}{2}\,\mathrm{Tr}\left[\left(\frac{\delta^2 S}{\delta\phi\delta\phi}+R_k\right)^{\!\!\!-\!1}\!\!\!\partial_t R_k\right].
\ee
One can also do slightly better and promote the couplings on the r.h.s to scale dependent ones to find a renormalization group improved flow. This is the approximation we will be content with in the present work. An interesting property of the flow equation \eqref{1loop_flow} is that promoting the actions on both sides to the full EAA will lead to an exact equation for this quantity \cite{wetterich_eqn, morris_erg} (see also \cite{btw} and references therein, especially \cite{tetradis_etal}). However, solving it requires resorting to suitable \nolinebreak truncation \nolinebreak schemes. 

The renormalization method adopted here can in principle be translated into more standard methods such as dimensional regularization and $\overline{\mathrm{MS}}$ \cite{scheme_dependence_frg}. However, eq.\eqref{1loop_flow} has the advantage of incorporating, in a straightforward way, the running of infinitely many couplings, which would otherwise require a resummation of an infinite number of diagrams. This is reflected in the fact that the beta functionals contain couplings in their denominators, as seen explicitly in eqs.(\ref{zeta_V}-\ref{zeta_P}).

Being just a matter of parametrization, the difference between linear and nonlinear models might sound irrelevant, given the general fact of quantum field theory that field redefinitions do not affect scattering amplitudes. However, beta functions are not physical quantities and are expected to depend on the choice of coordinates on the target space. One advantage of using the nonlinear parametrization is that the redundant couplings are all collected into a single function which is finally eliminated by a simple field redefinition. Apart from this, an important aspect of the approach taken here is the way the fluctuating fields are defined. In the standard linear theory it is customary, though not necessary, to adopt a linear splitting of the total field into background and fluctuations (including the possibility of a vanishing background field). In this work, instead, the fluctuations are parametrized nonlinearly via the exponential map. Although both of these choices for splitting the total field can be made regardless of the target space parametrization, and the two are physically equivalent \cite{pion_lagrangian}, they lead to inequivalent cutoff actions. In a theory with a linearly split field the cutoff action, while respecting the imposed symmetries, does not necessarily allow for possible enhanced symmetries in certain regions of the parameter space. In this sense the cutoff is not general enough. This is related to the fact that the cutoff term breaks general covariance. Instead, the background field method when accompanied by the exponential parametrization of fluctuations, allows for the most general choice of cutoff which respects all possible enhanced symmetries as it is invariant under general coordinate transformations. These points will become more clear in the subsequent sections.

\section{Set-up of the model}

We will consider a theory of four scalars in $d$ dimensional Euclidean space which respects electroweak symmetry. A suitable way to parametrize the field space is to use $\chi^\alpha$, $\alpha =1,2,3$, to assign an arbitrary parametrization to the three dimensional orbits of the symmetry group, which are homogeneous spaces isomorphic to $SU(2)_L\!\times\! U(1)_Y/U(1)_{em} = SU(2)$, and use the fourth field $\rho$ to label different orbits. In order to find the general form of the $SU(2)_L\!\times\! U(1)_Y$--invariant induced metric on the orbits, we make use of the Maurer-Cartan forms $L^I_\alpha$ which provide a dual basis for the left-invariant vector fields $L_I^\alpha$ of $SU(2)$. These are the generators of $SU(2)_R$ and therefore commute with $SU(2)_L$. In particular, $L_3^\alpha$ generates $U(1)_Y$ and commutes with $SU(2)_L$. The induced metric on the orbits then takes the general form
\be 
K(\rho) \,g_{\alpha\beta} + P(\rho)\,L^3_\alpha L^3_\beta,
\ee 
where $g_{\alpha\beta}$ is the metric invariant under $O(4)\supset SU(2)_L\!\times SU(2)_R$. With these considerations, at the second order of the derivative expansion, the dynamics is governed by the following Lagrangian
\be  \label{lagrangian_rhochi}
\mathcal{L} = \frac{1}{2}J(\rho)\partial_\mu \rho \partial^\mu \rho + V(\rho) + \frac{1}{2} K(\rho) \,\partial_\mu\chi^\alpha\partial^\mu\chi^\beta\,g_{\alpha\beta} + \frac{1}{2} P(\rho)\,\partial_\mu\chi^\alpha\partial^\mu\chi^\beta\,L^3_\alpha L^3_\beta.
\ee
This is in fact the Lagrangian in \eqref{lagrangian_hU} rewritten in terms of the fields $\rho$ and $\chi^\alpha$, except that the redundant function $J(\rho)$ is also included as it is generated under the renormalization group flow. Notice that, with abuse of notation, the same symbol has been used for the functions $K$ and $P$ despite the fact that they are now written in terms of $\rho$, which is related to the canonically normalized field $h$ through
\be \label{field_redef}
h(\rho) =\int_0^\rho \!\! d\sigma\;\sqrt{J(\sigma)}.
\ee
The metric $g_{\alpha\beta}$ and the one-forms $L^I_\alpha$ can be expressed in terms of the nonlinear sigma model field $U$ through the following relations
\be  
g_{\alpha\beta}= \frac{1}{2}\mathrm{Tr}\left(\partial_\alpha U^\dagger\partial_\beta U\right), \hspace{1cm} L^I_\alpha = \frac{i}{2}\mathrm{Tr}\left(U^\dagger\partial_\alpha U\sigma^I\right),
\ee
where $\sigma^I$ are the Pauli sigma matrices. Notice that we do not specify the parametrization of $U$ by $\chi^\alpha$. The first three terms in \eqref{lagrangian_rhochi} are invariant under $SU(2)_L\!\times SU(2)_R$, so after electroweak symmetry breaking, that is, picking a point in $\chi^\alpha$ space and expanding fields around this and the location of the minimum of the potential $\langle \rho\rangle$, the symmetry is broken to the vector subgroup $SU(2)_c$ known as custodial symmetry. The last term in \eqref{lagrangian_rhochi} breaks $SU(2)_L\!\times SU(2)_R$ explicitly to $SU(2)_L\!\times U(1)_Y$ and hence violates $SU(2)_c$ after electroweak symmetry is broken. In the Standard Model effective Lagrangian, custodial symmetry is broken starting from the dimension six (in $d=4$) 
operator $\vert H^\dagger\overset{{\hspace{0.5mm}\raisebox{-0.6mm}{\text{\scriptsize $\leftrightarrow$}}}}{\partial}_{\!\mu}\, H\vert^2$ (gauge fields neglected). 
This operator is included inside the last term in \eqref{lagrangian_hU}
\be  
h^4\left|\mathrm{Tr}\left(U^\dagger\partial_\mu U\sigma^3\right)\!\right|^2 = 4\, \vert H^\dagger\overset{{\hspace{0.5mm}\raisebox{-0.6mm}{\text{\scriptsize $\leftrightarrow$}}}}{\partial}_{\!\mu}\, H\vert^2,
\ee
where the explicit relation between the two complex fields $\phi^+$ and $\phi^0$ in the Higgs doublet $H$ and the four fields $h$ and $\chi^\alpha$ is given by the following equation
\be \label{hU_H_identification}
h\, U = \sqrt{2}\,(H^c \; H), \hspace{1cm} H^c = i\sigma^2 H^*, \hspace{1cm} H=\frac{1}{\sqrt{2}}\!\left(\!\!\ba{c}\phi^+ \\ \phi^0 \ea\!\!\!\right).
\ee  
In fact, the function $P(h)$ of \eqref{lagrangian_hU} evaluated at $\langle h \rangle$, where the potential takes its minimum, is related to the $\epsilon_1$ parameter \cite{ab_epsilon}, used in precision electroweak tests, through $P(\langle h\rangle)= - v^2\epsilon_1$, where $v$ is the weak scale. Recall that, by definition, the weak scale is also given by $v^2=K(\langle h\rangle)$. \\[1mm]
One can of course write the whole Lagrangian \eqref{lagrangian_rhochi} in terms of the Higgs doublet $H$, related to $\rho,\chi^\alpha$ through $\rho\, U=\sqrt{2}\,(H^c \; H)$, which differs from \eqref{hU_H_identification} in that $\phi^+$ and $\phi^0$ are not normalized canonically. This will give precisely the Lagrangian \eqref{lagrangian_H} if we make the identifications
\be 
J(\rho)=Z(\rho)+2\rho^2 Y(\rho), \hspace{4mm} K(\rho)=\rho^2Z(\rho), \hspace{4mm} P(\rho)=2\rho^4 T(\rho).
\ee
In the next section, we find the one-loop flow equations for the functions in \eqref{lagrangian_rhochi}. 

\section{Flow equations}  \label{flow_eqns}

We find it convenient to collect the fields $\rho$ and $\chi^\alpha$ into a four-component multiplet $\phi^i=(\rho,\chi^\alpha)$, with $i$ running over $0,1,2,3$, and rewrite the kinetic part of the theory in a manifestly reparametrization-invariant way  
\be \label{S_phi}
S =\int\!d^dx\,\left(\frac{1}{2}\tilde{G}_{ij}\partial_{\mu}\phi^{i}\partial^{\mu}\phi^{j}+V(\rho)\right),
\ee
where the metric $\tilde{G}_{ij}$ is equal to the $O(4)$ invariant metric $G_{ij}$ introduced in \cite{percacci_safari} plus a term proportional to $P(\rho)$ that breaks $O(4)$ explicitly to electroweak symmetry
\be  \label{metric}
\tilde{G}_{ij}= \left(
\ba{c|c} 
\!J(\rho) & \\ \hline & K(\rho)\,g_{\alpha\beta}+P(\rho)L^3_\alpha L^3_\beta
\ea 
\right).
\ee
This allows for a straightforward application of the methods developed in \cite{honerkamp1972,afm,boulware_brown,hps}, already employed for the renormalization group study of nonlinear sigma models in \cite{percacci_safari,codello_percacci,percacci_zanusso,fptz,fptv,bfptv,fzw}. This is despite the fact that we have already restricted to specific coordinate systems so that only reparametrizations that do not mix $\rho$ with $\chi^\alpha$ are allowed. In order to quantize the theory, it proves convenient to use the background field method and parametrize the fluctuations $\xi(x)$ around the background $\varphi(x)$ using the exponential map $\phi(x) = Exp_{\varphi(x)}\xi(x)$. In this way of splitting the total field, the fluctuations are vectors of the target manifold and transform linearly under any diffeomorphism. The symmetries are therefore preserved under quantization \cite{percacci_zanusso}. In order to write down the flow equation we need the piece in the EAA which is of second order in the fluctuations. This is given by the following expression
\be \label{sv} 
S^{(2)} = \frac{1}{2}\int \! d^{d}x \;\xi^i \big(-\!\tilde{\nabla}^2 \tilde{G}_{ij} +V'' \delta^{0}_{i} \delta^{0}_{j}-V' \tilde{\Gamma}^{0}_{ij}-\tilde{M}_{ij}\big) \xi^j, 
\ee
where $\tilde\nabla_\mu \xi^i = \partial_\mu \xi^i + \partial_\mu\varphi^k\tilde\Gamma^i_{kj}\xi^j$, $\tilde{M}_{ij} = \partial_{\mu}\varphi^m \partial^{\mu}\varphi^n \tilde{R}_{imjn}$ and the nonzero components of $\tilde{\Gamma}^{0}_{ij}$ are (see Appendix A)
\be  
\tilde\Gamma^{0}_{00} = \frac{J'}{2J}, \hspace{1cm}
\tilde\Gamma^0_{\alpha\beta} = - \frac{K'}{2J}\, g_{\alpha\beta} - \frac{P'}{2J}\, L^3_\alpha L^3_\beta.
\ee
The cutoff action being bilinear in the fluctuations takes the general form
\be  \label{cutoff_nl}
\Delta S_k = \frac{1}{2}\int\!d^dx\,\xi^i \left(\mathcal{R}_k\right)_{ij}\xi^j,
\ee
where the cutoff function $\mathcal{R}_k$ can depend on the background field. Here it is chosen to be proportional to the metric \eqref{metric}, with the optimized cutoff \cite{litim_optimized} used as the proportionality function 
\be \label{cutoff_function}
\left(\mathcal{R}_k\right)_{ij} = \tilde G_{ij}R_k, \hspace{1cm} R_k(z) = (k^2-z)\theta(k^2-z), \hspace{5mm} z \equiv -\tilde \nabla^2. 
\ee 
As pointed out in the Introduction, the cutoff \eqref{cutoff_nl} introduces a crucial difference between the covariant approach adopted here and the noncovariant approach where the fluctuating fields are defined as the difference between the total and background fields. To clarify this, let us recall that a symmetry that is imposed on the theory is the least amount of symmetry that we require, in general this can be enhanced at some regions of the parameter space. For instance, a model of four scalars with $O(4)$ symmetry can also become $O(5)$ symmetric, if the couplings are chosen appropriately. This extra symmetry, however, is not respected by the choice of cutoff in the noncovariant formulation. On the other hand, eq.\eqref{cutoff_nl} along with the leftmost relation in \eqref{cutoff_function}, because of its covariant nature, guarantees that any isometry of $\tilde G_{ij}$ will automatically be a symmetry of $\Delta S_k$, where the symmetry transformations act on the fluctuating fields only. 
\\
We write the sum of the second variation \eqref{sv} and the cutoff action \eqref{cutoff_nl} in the following way
\be  \label{msv} 
S^{(2)}+\Delta S_k = \frac{1}{2}\int \! d^{d}x \;\xi^i \big(\mathcal{P}_{ij} -\tilde{M}_{ij} \big) \xi^j,
\ee
where the Laplacian and the terms coming from the potential are collected into $\mathcal{P}_{ij}$ defined by
\be \label{pij}
\mathcal{P}_{ij} = \tilde{G}_{ij}P_k(z)+V''\delta^0_i\delta^0_j-V'\tilde{\Gamma}^0_{ij}, \hspace{1cm} P_k(z) = z+R_k(z).
\ee
We now have all the ingredients to find the one-loop flow equations. Using \eqref{msv} in the r.h.s of \eqref{1loop_flow} and expanding in $\tilde M_{ij}$ leads to   
\be \label{trace}
\frac{1}{2}\,\mathrm{Tr}\left[(\mathcal{P} -\tilde{M})^{\!-\!1}\dot{\mathcal{R}}_k\right] = \frac{1}{2}\,\mathrm{Tr}\left[\mathcal{P}^{-\!1}\dot{\mathcal{R}}_k\right]+\frac{1}{2}\,\mathrm{Tr}\left[\mathcal{P}^{-\!1}\tilde{M}\mathcal{P}^{-\!1}\dot{\mathcal{R}}_k\right]+\cdots,  
\ee 
where an overdot means derivation with respect to $t$. Notice that being interested in the one-loop beta functions, the metric $\tilde{G}_{ij}$, appearing implicitly inside the trace through $\mathcal{R}_k$, is considered to be scale independent so that $(\dot{\mathcal{R}}_k)_{ij} = \tilde{G}_{ij}\dot R_k$. The trace \eqref{trace}, when expanded in $\tilde M_{ij}$, gives, at zero order, the flow of the potential, and at first order, the flow of the functions $J$, $K$ and $P$. Appendix B gives the details of this computation. Let us define at this stage the relative $t$-derivatives of the functions $J$, $K$, $P$ and $V$, found by dividing the beta functionals by the corresponding functions
\be  \label{zetas}
\zeta_J = \frac{d}{dt}\log J, \hspace{1cm} \zeta_K = \frac{d}{dt}\log K, \hspace{1cm} \zeta_P = \frac{d}{dt}\log P, \hspace{1cm} \zeta_V = \frac{d}{dt}\log V.
\ee
These dimensionless zeta quantities are themselves functions of $J(\rho)$, $K(\rho)$, $P(\rho)$, $V(\rho)$ and their derivatives
. At this point we perform the field redefinition \eqref{field_redef} and write the zeta functions in terms of the dimensionless version $\tilde h$ of the canonically normalized field $h$ defined by $h = k^{\frac{d-2}{2}}\tilde h$. For this purpose we define the dimensionless functions denoted by a tilde $K(\rho)=k^{d-2}\tilde K(\tilde h)$, $P(\rho)=k^{d-2}\tilde P(\tilde h)$ and $V(\rho)=k^d\tilde V(\tilde h)$ and rewrite the zeta quantities in terms of these new functions. Doing this, all the dependence on $J(\rho)$ is absorbed into these new functions so that the zeta quantities will then depend on $\tilde K(\tilde h)$, $\tilde P(\tilde h)$ and $\tilde V(\tilde h)$ only, with no explicit $J$ dependence. The result of the computation is 
{\setlength\arraycolsep{3pt}
\bea
\zeta_V &=& c_d\left[\frac{1}{\tilde{V}(1+\tilde{V}'')} +\frac{2}{\tilde{V}(1+ \tilde{V}'\tilde{K}'/2\tilde{K})} +\frac{1}{\tilde{V}(1+\tilde{V}'(\tilde{K}+\tilde{P})'/2(\tilde{K}+\tilde{P}))}\right], \label{zeta_V} \\[5mm]
\zeta_J &=&  c_d \left[\frac{\tilde K'^2 -2\tilde K\tilde K''}{\tilde K^2(1+\tilde{V}'\tilde{K}'/2\tilde{K})^2} +\frac{(\tilde K+\tilde P)'^2 - 2(\tilde K+\tilde P)(\tilde K+\tilde P)''}{2(\tilde K+\tilde P)^2(1+\tilde{V}'(\tilde{K}+\tilde{P})'/2(\tilde{K}+\tilde{P}))^2}\right], \label{zeta_J} \\[5mm]
\zeta_K &=&  c_d \left[\frac{\tilde K'^2\!-\!2\tilde K\tilde K''}{2\tilde K^2(1\!+\!\tilde{V}'')^2} + \frac{4\tilde K\!-\! 12\tilde P\!-\!\tilde K'^2}{2\tilde K^2(1\!+\!\tilde{V}'\tilde{K}'/2\tilde{K})^2} \!+\! \frac{4(\tilde K\!+\!\tilde P)^2\!-\!\tilde K\tilde K'(\tilde K\!+\!\tilde P)'}{2\tilde K^2(\tilde K\!+\!\tilde P)(1\!+\!\tilde{V}'(\tilde{K}\!+\!\tilde{P})'/2(\tilde{K}\!+\!\tilde{P}))^2}\right], \label{zeta_K} \\[5mm]
\zeta_P &=&  c_d \left[\frac{\tilde K\tilde P'^2+2\tilde K\tilde K'\tilde P'-2\tilde K(\tilde K+\tilde P)\tilde P''-\tilde P\tilde K'^2}{2\tilde K\tilde P(\tilde K+\tilde P)(1+\tilde{V}'')^2} + \frac{28\tilde K\tilde P+4\tilde K^2+8\tilde P^2-2\tilde K\tilde K'\tilde P'-\tilde K\tilde K'^2}{2\tilde K^2\tilde P(1+ \tilde{V}'\tilde{K}'/2\tilde{K})^2} \right. \nn\\
&& \left. \;\; +\, \frac{\tilde K\tilde K'(\tilde K+\tilde P)'-4(\tilde K+\tilde P)^2}{2\tilde K\tilde P(\tilde K+\tilde P)(1+\tilde{V}'(\tilde{K}+\tilde{P})'/2(\tilde{K}+\tilde{P}))^2}\right],  \label{zeta_P}
\eea}%
where $1/c_d= (4\pi)^{d/2}\Gamma(d/2+1)$. The derivatives on the tilde functions $\tilde V$ and $\tilde P$ are taken with respect to $\tilde h$. The r.h.s expressions are written in terms of the tilde functions and therefore there is no $J$ appearing explicitly on the r.h.s. In fact from the definitions of the tilde functions and \eqref{field_redef} it can be shown that $\tilde V' = V'/\sqrt{J}$ and $\tilde V'' = V''/J-V'J'/2J^2$ with similar relations for $\tilde K$ and $\tilde P$. This gives the relations through which the $J$ function implicitly appears. In the special case where the potential takes a constant value, the first equation \eqref{zeta_V} above gives the flow of this constant $dV/dt=4c_d$ and the last three equations (\ref{zeta_J},\ref{zeta_K},\ref{zeta_P}) reduce to a Ricci flow \cite{friedan_nl, codello_percacci}
\be 
\frac{d\tilde G_{ij}}{dt} = 2c_d k^{d-2} \tilde R_{ij}.  
\ee
This can be checked using the expressions for the Ricci tensor (\ref{Ricci00}) and (\ref{Ricciab}) given in Appendix A. The redundant function $J$ still has a flow of its own, but this is not of interest to us because it is absorbed into the functions $\tilde K(\tilde h)$, $\tilde P(\tilde h)$ and $\tilde V(\tilde h)$. The flow of these three functions can be written using the zeta quantities as 
{\setlength\arraycolsep{3pt}
\bea
\frac{\partial\tilde V}{\partial t} &=&
\left(\zeta_V-d\right)\tilde{V}
+\frac{d-2}{2}\,\tilde{h}\,\tilde{V}' -\frac{1}{2}\,\tilde{V}'\!\int_{0}^{\tilde{h}}\!\!\!\! d\sigma \, \zeta_J (\sigma),
\label{flowV} \\[1.2mm]
\frac{\partial\tilde{K}}{\partial t} &=&
\left(\zeta_K -d+2\right)\tilde{K}
+\frac{d-2}{2}\,\tilde{h}\,\tilde{K}' -\frac{1}{2}\,\tilde{K}'\!\int_{0}^{\tilde{h}}\!\!\!\! d\sigma \, \zeta_J (\sigma),
\label{flowK} \\[1.2mm]
\frac{\partial\tilde{P}}{\partial t} &=&
\left(\zeta_P -d+2\right)\tilde{P}
+\frac{d-2}{2}\,\tilde{h}\,\tilde{P}' -\frac{1}{2}\,\tilde{P}'\!\int_{0}^{\tilde{h}}\!\!\!\! d\sigma \, \zeta_J (\sigma),
\label{flowP}
\eea}%
where the $t$-derivatives of $\tilde K(\tilde h)$, $\tilde P(\tilde h)$ and $\tilde V(\tilde h)$ are taken keeping the field $\tilde h$ fixed. These are found using eqs.\eqref{zetas} and the scale dependence of $\tilde h$ \eqref{field_redef} which leads to  
\be
\frac{d\tilde h}{dt}
=\frac{2-d}{2}\,\tilde h+\frac{1}{2}\int_0^{\tilde h} \!\! d\sigma\;\zeta_J(\sigma).
\ee
In the following sections we analyse three solutions of the above flow equations (\ref{flowV}-\ref{flowP}) with geometries of flat space $\mathbb{R}^4$, cylinder $\mathbb{R}\times S^3$ and sphere $S^4$, and allow for electroweak-preserving fluctuations around these solutions. We concentrate on the fluctuations that do not respect $O(4)$ symmetry, and find the eigenperturbations and their corresponding eigenspectrum.

\section{The Gaussian fixed point (flat geometry)} 

The flat metric is given by the choices $\tilde K = \tilde h^2$ and $\tilde P = 0$. In this case there will be no running in the two functions $\tilde K$ and $\tilde P$, i.e. $\partial\tilde K/\partial t = 0$ and $\partial\tilde P/\partial t = 0$, and the potential $\tilde V$ will have a flow of the following form 
\be \label{flowVflat}
\frac{\partial\tilde V}{\partial t} = c_d\left[\frac{1}{1+\tilde V''}+\frac{3h}{h+\tilde V'}\right]-d\tilde V.
\ee
If we further restrict to constant values of the potential, this will lead to a free theory whose structure is preserved under the renormalization group with only a running constant potential. In other words, for this choice of couplings, the symmetry is enhanced to rotation and translation in the four-dimensional field space. The constant $\tilde{V}_* =4c_d/d$ is clearly a fixed point of the flow \eqref{flowVflat}, which together with $\tilde K_* = \tilde h^2$ and $\tilde P_* = 0$ specifies the Gaussian fixed point. The next information we can easily extract from the flow equations is the eigen-perturbations around the fixed point and their corresponding eigenvalues. These are found using the linearized form of the flow equations (\ref{flowV}-\ref{flowP}) around the Gaussian fixed point
{\setlength\arraycolsep{3pt}
\bea
\lambda\delta\tilde V  &=&
\delta\zeta_V\,\tilde{V}_*
+\frac{d-2}{2}\,\tilde{h}\,\delta\tilde{V}',
\label{flowV_flat_1} \\[0.7mm]
\lambda\delta\tilde{K} \hspace{-0.3mm} &=&
\delta\zeta_K\,\tilde{K}_* -(d-2)\delta\tilde{K}
+\frac{d-2}{2}\,\tilde{h}\,\delta\tilde{K}'-\frac{1}{2}\,\tilde{K}'_*\!\int_{0}^{\tilde{h}}\!\!\!\! d\sigma \; \delta\zeta_J(\sigma),
\label{flowK_flat_1} \\[1.5mm]
\lambda\delta\tilde{P}  &=&
\delta(\zeta_P\tilde{P}) -(d-2)\delta\tilde{P}
+\frac{d-2}{2}\,\tilde{h}\,\delta\tilde{P}',  
\label{flowP_flat_1} 
\eea}%
where $\delta\zeta_K$, $\delta\zeta_J$, $\delta\zeta_V$ and $\delta(\zeta_P\tilde{P})$ are respectively the first order values of $\zeta_K$, $\zeta_J$, $\zeta_V$ and $\zeta_P\tilde{P}$ in the variations $\delta\tilde{K}$, $\delta\tilde{V}$ and $\delta\tilde{P}$ in an expansion around the fixed point. These are given explicitly by the following expressions
{\setlength\arraycolsep{3pt}
\bea
\delta\zeta_V &=& -\frac{d(d\,\tilde h\,\delta\tilde V+3c_d\delta\tilde V'+c_d\tilde h\delta\tilde V'')}{4\tilde h c_d}, \label{delta_zeta_V_flat} \\
\delta\zeta_J\, &=& c_d\,\frac{-6\delta\tilde K+6\tilde h \delta\tilde K'-3\tilde h^2\delta\tilde K''-2\delta\tilde P+2\tilde h \delta\tilde P'-\tilde h^2\delta\tilde P''}{\tilde h^4}, \label{delta_zeta_J_flat} \\
\delta\zeta_K &=& c_d\,\frac{2\delta\tilde K -2\tilde h\delta\tilde K'-\tilde h^2\delta\tilde K''- 2\delta\tilde P-\tilde h \delta\tilde P'}{\tilde h^4}, \label{delta_zeta_K_flat} \\
&& \delta(\zeta_P\tilde{P}) = c_d\,\frac{8\delta\tilde P + \tilde h \delta \tilde P'-\tilde h^2 \delta \tilde P''}{\tilde h^2}. \label{delta_zeta_PP_flat}
\eea}%
The first equation \eqref{flowK_flat_1} is of integro-differential type. In order to bring it into pure differential form one can divide it by $\tilde{K}_*$ 
\be  
\lambda\Delta\tilde K =
\frac{\tilde{h}}{2}\delta\zeta_K -\frac{d-2}{2}\Delta\tilde K
+\frac{d-2}{2}\,\tilde{h}\Delta\tilde K'-\frac{1}{2}\int_{0}^{\tilde{h}}\!\!\!\! d\sigma \; \delta\zeta_J(\sigma), \hspace{1cm} \Delta\tilde K \equiv \frac{\delta\tilde K}{\tilde K'_*}, 
\label{flowK_flat_intdiff} 
\ee
and then differentiate with respect to $\tilde{h}$ to get
\be 
\lambda\,\mathcal{K} =
\frac{1}{2}\frac{d(\tilde{h}\delta\zeta_K)}{d\tilde{h}}+\frac{d-2}{2}\,\tilde{h}\,\mathcal{K}'-\frac{1}{2}\,\delta\zeta_J, \hspace{1cm} \mathcal{K} \equiv \Delta\tilde K'.
\label{flow_calK_flat} 
\ee
Of course $\delta\zeta_K$ and $\delta\zeta_J$ should now be written in terms of $\mathcal{K}$. It is easily seen that on substituting $\delta\tilde K$ in terms of $\Delta\tilde K$ in the expression \eqref{delta_zeta_K_flat}, $\Delta\tilde K$ does not appear undifferentiated, so equation \eqref{flow_calK_flat} together with \eqref{flowP_flat_1} gives rise to a set of two coupled second-order differential equations. Now, having found $\mathcal{K}$, the function $\delta K$ is specified, up to an integration constant times $2\tilde{h}$, by
\be \label{delta_K_flat}
\delta\tilde K(\tilde{h}) = 2\tilde{h}\int_0^{\tilde{h}} \!\!\! d\sigma \;\mathcal{K}(\sigma).
\ee 
This integration constant vanishes by the requirement that $\delta\tilde K(\tilde{h})$ satisfy \eqref{flowK_flat_intdiff}. This is because eq.\eqref{flowK_flat_intdiff}, when differentiated, is satisfied by the solution \eqref{delta_K_flat}, so the undifferentiated version \eqref{flowK_flat_intdiff} is satisfied up to a constant. On the other hand, from the following analysis, the solutions $\mathcal{K}$ and $\delta\tilde P$ will turn out to be even functions. This, together with the solution \eqref{delta_K_flat} implies that all the terms in \eqref{flowK_flat_intdiff} are odd, so the constant must vanish. \\[1mm]
Using the expressions (\ref{delta_zeta_V_flat}-\ref{delta_zeta_PP_flat}) the explicit form of the linearized equations \eqref{flowV_flat_1}, \eqref{flowP_flat_1} and \eqref{flow_calK_flat} is
{\setlength\arraycolsep{3pt}
\bea
0 &=& \delta\tilde V'' \!+\left[\frac{(2-d)h}{2c_d}+\frac{3}{h}\right]\delta\tilde V'+ \frac{d+\lambda}{c_d}\,\delta\tilde V,  \label{eqn_flat_V}
\\[3mm]
0 &=& \,\mathcal{K}'' \,+\left[\frac{(2-d)h}{2c_d}+\frac{1}{h}\right]\mathcal{K}'+ \left[\frac{\lambda}{c_d}-\frac{4}{h^2}\right]\mathcal{K} +\frac{1}{h^3}\,\delta\tilde P'-\frac{4}{h^4}\,\delta\tilde P, \label{eqn_flat_K} \\[3mm]
0 &=& \delta\tilde P''\!+\left[\frac{(2-d)h}{2c_d}-\frac{1}{h}\right]\delta\tilde P'+\left[\frac{d+\lambda -2}{c_d}-\frac{8}{h^2}\right]\delta\tilde P.  \label{eqn_flat_P}
\eea}%
For $\delta\tilde P=0$ these equations reproduce the results of \cite{percacci_safari} at $N=4$, where the solutions  to the two decoupled equations were found to be  
{\setlength\arraycolsep{4pt}
\bea
\delta\tilde{V}_i &=& \! \phantom{}_{1\!}F_1(-i,2,\bar h^2), \hspace{1.04cm} \lambda^V_i = -d+(d-2)i, \hspace{5mm} i=0,1,2,\ldots, \hspace{5mm} \bar h \equiv \sqrt{\frac{d-2}{4c_d}}\tilde h  \label{sol_flat_V}
\\ 
\mathcal{K}^{\mathrm{hom}}_i \!\!\! &=& \bar h^2\,\phantom{}_{1\!}F_1(-i,3,\bar h^2), \hspace{5mm} \lambda^K_i = (d-2)(i+1), \hspace{5mm} i=0,1,2,\ldots  \label{sol_flat_K_hom}
\eea}%
Here instead we are interested in the solutions with nonzero $\delta\tilde P$, and therefore $\mathcal{K}^{\mathrm{hom}}_i$ is the solution to the homogeneous part of \eqref{eqn_flat_K} only. So we first have to solve eq.\eqref{eqn_flat_P}. The solution is
\be \label{sol_flat_P}
\delta\tilde{P}_i = \bar h^4\,\phantom{}_{1\!}F_1(-i,4,\bar h^2), \hspace{1cm} \lambda^P_i = (d-2)(i+1), \hspace{5mm} i=0,1,2,\ldots.
\ee
In this case, we can also have nonzero $\delta\tilde V$ if for some $i,j$ we have $\lambda^P_i=\lambda^V_j$. This happens for example in $d=3,4$ where $\lambda^P_i=\lambda^V_{i+4}$ and $\lambda^P_i=\lambda^V_{i+3}$ respectively. Now lets come to eq.\eqref{eqn_flat_K}. The solution to its homogeneous version is given by \eqref{sol_flat_K_hom}, and it has the same eigenvalue as \eqref{sol_flat_P}. So the general solution to \eqref{eqn_flat_K} is given by an arbitrary coefficient of \eqref{sol_flat_K_hom} plus any function that solves \eqref{eqn_flat_K}. To find this specific solution we write equations \eqref{eqn_flat_P} and \eqref{eqn_flat_K} in the following compact form 
{\setlength\arraycolsep{2pt}
\bea 
\mathfrak{L}_{K\!K}\,\mathcal{K} + \mathfrak{L}_{K\!P}\;\delta\tilde P &=& \lambda \;\mathcal{K}, \label{comp_eq_Flat_K} \\
\mathfrak{L}_{P\!P}\;\delta\tilde P &=& \lambda \,\delta\tilde P. \label{comp_eq_Flat_P}
\eea}%
In order to treat the functions $\delta P$ and $\delta K$ on the same footing, we define the operator $D$ as
\be  
D \equiv \frac{d}{d\bar h}\,\frac{1}{2\bar h},
\ee
and using $\mathcal{K} =D \delta\tilde K$, we rewrite eq.\eqref{comp_eq_Flat_K} in terms of $\delta\tilde K$, and take $\delta\tilde P=\delta\tilde P_i$ and $\lambda = \lambda^{\!P}_i$ to make sure eq.\eqref{comp_eq_Flat_P} is satisfied
\be 
\mathfrak{L}_{K\!K}D\,\delta\tilde K + \mathfrak{L}_{K\!P}\,\delta\tilde P_i = \lambda^{\!P}_i D\,\delta\tilde K. \label{comp_eq_Flat_K_1} 
\ee
It can be verified that the following relation holds between the differential operators $\mathfrak{L}_{K\!K}$, $\mathfrak{L}_{P\!P}$, $\mathfrak{L}_{K\!P}$ and $D$
\be  
\mathfrak{L}_{K\!K}D-D\,\mathfrak{L}_{P\!P}=3\mathfrak{L}_{K\!P}.
\ee
Using this identity, we can write eq.\eqref{comp_eq_Flat_K_1} in the following way
\be
D(\mathfrak{L}_{P\!P}-\lambda^{\!P}_i)\delta\tilde K + \mathfrak{L}_{K\!P}(3\,\delta\tilde K +\delta\tilde P_i) = 0. \label{comp_eq_Flat_K_2}
\ee
Choosing $\delta \tilde K$ to be a solution to \eqref{comp_eq_Flat_P} with $\lambda = \lambda^{\!P}_i$ makes the first term vanish. The second term suggests the proportionality factor. So from eq.\eqref{comp_eq_Flat_K_2} it is clear that $\delta\tilde K = -\delta \tilde P_i/3$ solves the equation. The general eigensolution of eqs.(\ref{eqn_flat_V},\ref{eqn_flat_K},\ref{eqn_flat_P}) with $\delta\tilde P\neq 0$ will then be
\be  
\delta\tilde{P}_i = \bar h^4\,\phantom{}_{1\!}F_1(-i,4,\bar h^2), \hspace{8mm}
\delta\tilde{K}_i = C\,\delta\tilde{K}^{\mathrm{hom}}_i-\frac{1}{3}\,\delta\tilde P_i, \hspace{8mm} 
\lambda_i = (d-2)(i+1), \hspace{3mm} i=0,1,2,\ldots,
\ee
where $\delta\tilde{K}^{\mathrm{hom}}_i$ is given by the same expression as \eqref{delta_K_flat} with $\mathcal{K}$ replaced by $\mathcal{K}^{\mathrm{hom}}_i$. More explicitly
{\setlength\arraycolsep{2pt}
\be \label{result_flat}
\boxed{
\ba{lll}  
\delta\tilde{K}_i &=& \displaystyle C_i\,\bar h^2\,\phantom{}_{1\!}F_1(-i,3,\bar h^2)-\frac{1}{3}\,\bar h^4\,\phantom{}_{1\!}F_1(-i,4,\bar h^2) \\
\delta\tilde{P}_i &=& \bar h^4\,\phantom{}_{1\!}F_1(-i,4,\bar h^2)  
\ea
\hspace{1cm} 
\lambda_i = (d-2)(i+1), \hspace{5mm} i=0,1,2,\ldots\,}
\ee}%
As mentioned before, in $d=3,4$ we can also have a non vanishing eigenperturbation in the potential, proportional  to $\delta\tilde V_{i+4}$ and $\delta\tilde V_{i+3}$ respectively. There is no such possibility in higher dimensions. \\[1mm]
In fact the eigensolutions \eqref{result_flat} fulfil our expectations regarding the $O(4)$-violating perturbations around a Gaussian fixed point. The spectrum gives simply the dimensions of the couplings of the higher-dimensional operators $(H^\dagger H)^i\vert H^\dagger\overset{{\hspace{0.5mm}\raisebox{-0.6mm}{\text{\scriptsize $\leftrightarrow$}}}}{\partial}_{\!\mu}\, H\vert^2$ which violate custodial symmetry, and $i$ is the highest power of $H^\dagger H$ multiplying $\vert H^\dagger\overset{{\hspace{0.5mm}\raisebox{-0.6mm}{\text{\scriptsize $\leftrightarrow$}}}}{\partial}_{\!\mu}\, H\vert^2$ in the eigenperturbation $\delta\tilde{P}_i$.

\section{Cylindrical geometry}  

Choosing the functions $\tilde K$ and $\tilde P$ to be constant, the Goldstone sector decouples from the Higgs and it will consequently remain so under the renormalization group flow. We parametrize the constants $\tilde K$ and $\tilde P$ by the two dimensionless parameters $\tilde f$ and $a$ as $\tilde K = 1/\tilde f^2$ and $\tilde P = -a/\tilde f^2$, and find the following expressions for the zeta quantities: $\zeta_J= 0$, $\zeta_K= 4c_d(1+a)\tilde f^2$ and $\zeta_P= 4c_d(3-a)\tilde f^2$. Inserting these expressions into the flow equations (\ref{flowK},\ref{flowP}) gives
{\setlength\arraycolsep{3pt}
\bea \label{flowKcyl}
\frac{d}{dt}\left(\frac{1}{\tilde f^2}\right)  &=& \frac{2-d}{\tilde f^2}+4c_d(a+1), \\ \label{flowPcyl}
-\frac{d}{dt}\left(\frac{a}{\tilde f^2}\right) &=& \frac{a(d-2)}{\tilde f^2}+4c_d\,a(a-3),
\eea}%
from which the flow equations for $\tilde f^2$ and $a$ follow
\be \label{flowfa}
\frac{d\tilde f^2}{dt} = (d-2)\tilde f^2 - 4c_d(1+a)\tilde f^4, \hspace{1cm}
\frac{da}{dt} = 8c_da(1-a)\tilde f^2.
\ee
In fact the first terms in (\ref{flowKcyl},\ref{flowPcyl}) come from the canonical dimensions and the second terms come from the term $2c_d R_{\alpha\beta}$, proportional to the Ricci tensor, which reduces in this case to the simple form (see eq.\eqref{Ricciab} in Appendix A)
\be  \label{Ricci_cyl}
\tilde R_{\alpha\beta} = 2\left(1+a\right) g_{\alpha\beta} + 2a(a-3) L^3_\alpha L^3_\beta.
\ee 
This is because we are working in the one-loop approximation and because the flow of the potential has decoupled from that of $\tilde f$ and $a$. The flow of the potential $\tilde V$, which is independent of $\tilde f^2$ and $a$, is
\be \label{flowVcyl}
\frac{\partial\tilde V}{\partial t} = c_d\left[\frac{1}{1+\tilde V''}+3\right] - d\, \tilde V.
\ee 
A constant potential therefore remains constant, with the flow $\partial_t\tilde V=4c_d-d\tilde V$. The constant value $\tilde V_* = 4c_d/d$ is a fixed point of the flow equation \eqref{flowVcyl} in any space-time dimension, although in principle other fixed points like the Wilson-Fisher in $d=3$ exist. When all three functions are constant (not necessarily at the fixed point), the symmetry is enhanced from electroweak to electroweak plus shift invariance of $h$.\\[1mm]
For the two quantities $\tilde f$ and $a$, three fixed points can be identified from \eqref{flowfa}. The first one, at which $\tilde f$ vanishes and $a$ is left arbitrary, gives the trivial fixed point. At the second fixed point, $a$ vanishes and $\tilde f$ takes the value $\tilde f^2=(d-2)/4c_d$. This is the fixed point with cylindrical geometry where the symmetry is enhanced to rotations, and translations along the cylinder axis, and the one we will be finally dealing with in this section. A third fixed point, given by $a=1$ and $\tilde f^2=(d-2)/8c_d$, can be identified from \eqref{flowfa}, but this is not a fixed point of the full flow equations (\ref{flowK}-\ref{flowP}) because $\tilde K_*+\tilde P_*=0$ and so terms like $(\tilde K'+\tilde P')/(\tilde K+\tilde P)$ which appear inside the zetas (\ref{zeta_V}-\ref{zeta_P}) will not be well defined anymore. \\[1mm]
Let us now discuss the linearized equations around a generic solution to the flows of the three constants $\tilde V$, $\tilde K$, $\tilde P$. Linearization of eqs.(\ref{flowV}-\ref{flowP}) in this case leads to  
{\setlength\arraycolsep{3pt}
\bea
\lambda\delta\tilde V  &=&
\delta\zeta_V\,\tilde{V} +\frac{d-2}{2}\,\tilde h\,\delta V',
\label{flowV_cyl_1} \\[1mm]
\lambda\delta\tilde{K}\hspace{-0.3mm} &=&
\delta\zeta_K\,\tilde{K} +\frac{d-2}{2}\,\tilde h\,\delta\tilde K',
\label{flowK_cyl_1} \\[1mm]
\lambda\delta\tilde{P} &=&
\delta(\zeta_P\tilde{P}) -(d-2)\delta\tilde{P}+\frac{d-2}{2}\,\tilde h\,\delta P',
\label{flowP_cyl_1} 
\eea}%
where, parametrizing also the potential as $\tilde V=4c_d/b$, for which the fixed point occurs at $b=d$, the quantities $\delta\zeta_V$, $\delta\zeta_J$, $\delta\zeta_K$ and $\delta(\tilde P\zeta_P)$ are expressed in terms of the linear fluctuations as
{\setlength\arraycolsep{3pt}
\bea
\delta\zeta_V   &=& -\frac{b\,(b\,\delta\tilde V+c_d\,\delta\tilde V'')}{4c_d}, \label{delta_zeta_V_cyl} \\[0.9mm]
\delta\zeta_J\, &=& -\frac{c_d \tilde f^2((3-2a)\delta\tilde K''+\delta\tilde P'')}{1-a}, \label{delta_zeta_J_cyl} \\[2.2mm]
\delta\zeta_K   &=& -c_d\tilde f^2\big(4(1+2a)\tilde f^2\delta\tilde K + \delta\tilde K'' + 4\tilde f^2\delta\tilde P\big), \label{delta_zeta_K_cyl} \\[3.1mm]
&& \hspace{-1.4cm} \delta(\zeta_P\tilde{P}) \, = \, -c_d\big(4a(2a-3)\tilde f^2 \delta\tilde K + 4(2a-3)\tilde f^2 \delta\tilde P +\delta\tilde P''\big). \label{delta_zeta_PP_cyl}
\eea}%
Inserting the above expressions into (\ref{flowV_cyl_1}-\ref{flowP_cyl_1}) gives the explicit form of the linearized equations 
{\setlength\arraycolsep{3pt}
\bea
0 &=& \delta\tilde V'' -\frac{(d-2)h}{2c_d}\,\delta\tilde V'+ \frac{\lambda +b}{c_d}\,\delta\tilde V, \label{eqn_cyl_V} \\[2mm]
0 &=& \delta\tilde K'' -\frac{(d-2)h}{2c_d}\,\delta\tilde K'+ \frac{\lambda + 4c_d\tilde f^2(1+2a)}{c_d}\,\delta\tilde K +4\tilde f^2\,\delta\tilde P,  \label{eqn_cyl_K} \\[2mm]
0 &=& \delta\tilde P''-\frac{(d-2)h}{2c_d}\,\delta\tilde P'+\frac{\lambda +d-2-4c_d\tilde f^2(3-2a)}{c_d}\,\delta\tilde P-4a(3-2a)\tilde f^2\,\delta\tilde K. \label{eqn_cyl_P}
\eea}%
The first equation is decoupled from the last two and admits the set of solutions
\be 
\delta\tilde{V}_i = \phantom{}_{1\!}F_1(-i,1/2,\bar h^2), \hspace{7mm} \lambda^V_i =(d-2)i-b, \hspace{5mm} i=0,1,2,\ldots, \hspace{5mm} \bar h \equiv \sqrt{\frac{d-2}{4c_d}}\tilde h.  \label{sol_cyl_V}
\ee
The two equations for $\delta\tilde K$ and $\delta\tilde P$ have to be solved simultaneously. To solve them we first find the solutions to the homogeneous versions: $\delta\tilde P=0$ in \eqref{eqn_cyl_K} and $\delta\tilde K=0$ in \eqref{eqn_cyl_P}. These equations are the same as eq.\eqref{eqn_cyl_V} with $\lambda$ shifted appropriately. So the solutions are 
{\setlength\arraycolsep{3pt}
\bea 
\delta\tilde{P}^{\,\mathrm{hom}}_i\! &=& \phantom{}_{1\!}F_1(-i,1/2,\bar h^2), \hspace{8mm} \lambda^P_i = (d-2)(i-1)+4c_d\tilde f^2(3-2a), \hspace{5mm} i=0,1,2,\ldots  \label{sol_cyl_P_hom} \\
\delta\tilde{K}^{\,\mathrm{hom}}_i  \!\! &=& \phantom{}_{1\!}F_1(-i,1/2,\bar h^2), \hspace{8mm} \lambda^K_i \hspace{-0.4mm} = (d-2)i-4c_d\tilde f^2(1+2a), \hspace{5mm} i=0,1,2,\ldots  \label{sol_cyl_K_hom}
\eea}%
Now, the coupled system of equations (\ref{eqn_cyl_K},\ref{eqn_cyl_P}) is solved by plugging in the ansatz $c_K\delta\tilde{K}^{\,\mathrm{hom}}_i$ and $c_P\delta\tilde{P}^{\,\mathrm{hom}}_i$ which leads to an algebraic eigenvalue problem for the eigenvector $(c_K,c_P)$ and eigenvalue $\lambda$. This is easily solved to give 
{\setlength\arraycolsep{3pt}
\bea 
\delta\tilde{K}^\pm_i\!\! &=& \frac{\lambda^K_i-\lambda^P_i\pm\sqrt{(\lambda^K_i-\lambda^P_i)^2-64c^2_d\tilde f^4a(3-2a)}}{8c_d\tilde f^2a(3-2a)}\delta\tilde{K}^{\,\mathrm{hom}}_i, \hspace{1cm} \delta\tilde{P}^\pm_i = \delta\tilde{P}^{\,\mathrm{hom}}_i, \label{eigenpert_cyl} \\
\lambda^\pm_i &=& \frac{\lambda^K_i+\lambda^P_i\pm\sqrt{(\lambda^K_i-\lambda^P_i)^2-64c^2_d\tilde f^4a(3-2a)}}{2}, \hspace{1cm} i=0,1,2,\ldots \label{eigenval_cyl}
\eea}%
Up to now we have tried to keep the analysis as general as possible. The generic solution to the flow equations around which we have linearized describes a cylinder with a squashed sphere base whose shape is deforming through ``renormalization group time'' $t$. From now on we specialize to the case $a=0$ which corresponds to a cylinder (with spherical base) which is expanding or contracting, say, as we move towards the UV, depending on whether the radius is smaller or bigger than the fixed point value, as can be seen from the $\tilde f$ beta function in \eqref{flowfa}. In this limit, one of the eigenperturbations in \eqref{eigenpert_cyl} preserves $O(4)$ while the other one, in which we are interested \nolinebreak[4] simplifies \nolinebreak to
\be \label{eigensol_cyl_a=0}
\delta\tilde{K}_i\! = -\frac{4c_d\tilde f^2}{16c_d\tilde f^2-d+2}\delta\tilde{K}^{\,\mathrm{hom}}_i\!\!\!\!, \hspace{6mm} \delta\tilde{P}_i = \delta\tilde{P}^{\,\mathrm{hom}}_i\!\!\!\!, \hspace{6mm} \lambda_i = (d-2)(i-1)+12c_d\tilde f^2, \hspace{4mm} i=0,1,2,\ldots 
\ee
From the expression for the eigenvalues \eqref{eigensol_cyl_a=0} it is seen that when the cylinder radius $\tilde f^{-\!1}$ is small enough, or explicitly when $\tilde f^2>(d-2)/12c_d$, which also includes the fixed point value, the eigenvalues are all positive and therefore the perturbations are IR irrelevant, while for $\tilde f^2<(d-2)/12c_d$ the lowest-order perturbation grows in the IR.\\[1mm]
In this $a=0$ case, the last term in \eqref{eqn_cyl_P} vanishes, so the above eigenfunction might not be the unique one and one can add to $\delta\tilde{K}_i$ any solution of the homogeneous version of \eqref{eqn_cyl_P} with $\lambda=\lambda_i$. Such a solution exists if $\lambda_i=\lambda^K_j$, 
for some $j$. This happens for example at the fixed point $\tilde f^2_* = (d-2)/4c_d$ for $j=i+3$, so that one can add $\delta\tilde{K}^{\,\mathrm{hom}}_{i+3}$, with arbitrary coefficient, to the solution \eqref{eigensol_cyl_a=0}. In summary, at the fixed point the eigensolutions with nonzero $\delta\tilde P$ are found to be
{\setlength\arraycolsep{2pt}
\be 
\boxed{
\ba{lll}
\delta\tilde{K}_i &=& \displaystyle C_i\,\phantom{}_{1\!}F_1(-i-3,1/2,\bar h^2)-\frac{1}{3}\,\phantom{}_{1\!}F_1(-i,1/2,\bar h^2) \\[0.8mm]
\delta\tilde{P}_i &=& \phantom{}_{1\!}F_1(-i,1/2,\bar h^2)
\ea \hspace{8mm} 
\lambda_i = (d-2)(i+2), \hspace{5mm} i=0,1,2,\ldots}
\ee}%
Finally, we would like to know if in the presence of nonzero $O(4)$-violating eigenperturbations $\delta\tilde P$ around the fixed point we can also have nonzero eigenperturbations $\delta\tilde V$ in the potential. To find out, we need to see if they can have the same eigenvalues, which means if there are $i$ and $j$ such that $\lambda^P_i=\lambda^V_j$. A simple analysis shows that $\lambda^P_i=\lambda^V_{i+5}$ in $d=3$ and $\lambda^P_i=\lambda^V_{i+4}$ in $d=4$, so that one can also have an eigenperturbation in the potential, proportional to $\delta\tilde V_{i+5}$ and $\delta\tilde V_{i+4}$ respectively.

\section{Spherical geometry} 

Finally, let us restrict our two functions $\tilde K$ and $\tilde P$ to $\tilde K = \tilde f^2\sin^2(\tilde h/\tilde f)$ and $\tilde P =0$, where $\tilde f$ is a dimensionless parameter. This gives the spherical geometry for which the symmetry is enhanced to $O(5)$ when also accompanied by the choice of constant $\tilde V$. Because of this extra symmetry the structure of the Lagrangian will be preserved and they are only the constants $\tilde f$ and $\tilde V$ that run under the renormalization group flow: $d\tilde f/dt=(6c_d-(d-2)\tilde f^2)/2\tilde f$, $ d\tilde V/dt=4c_d-d\tilde V$. The fixed point value of the sphere radius and the potential is thus given by $\tilde f^2_*=6c_d/(d-2)$, $\tilde V_* =4c_d/d$. We linearize the flow equations (\ref{flowV}-\ref{flowP}) around this $O(5)$ symmetric geometry with $\tilde f$ and $\tilde V\equiv 4c_d/b=\mathrm{const.}$ satisfying the above flow equations. Just like the cylindrical case, this solution describes an expanding (when $\tilde f<\tilde f_*$) or contracting (when $\tilde f>\tilde f_*$) four--sphere, as we move towards the UV, ending up at the fixed point. The linearized equations are
{\setlength\arraycolsep{3pt}
\bea
\lambda\,\delta\tilde V &=&
\delta\zeta_V\,\tilde{V},
\label{flowV_sph_1} \\[0.8mm]
\lambda\,\delta\tilde{K}\hspace{-0.4mm} &=&
\delta\zeta_K\,\tilde{K} -\frac{1}{2}\tilde{K}'\int_{0}^{\tilde{h}}\!\!\!\! d\sigma \; \delta\zeta_J(\sigma),
\label{flowK_sph_1} \\[1.2mm]
\lambda\,\delta\tilde{P} &=&
\delta(\zeta_P\tilde{P}) -(d-2)\delta\tilde{P}.
\label{flowP_sph_1}
\eea}%
For convenience we define $\bar h$ by $\tilde h \equiv \tilde f\bar h$ and the barred functions by $\delta\tilde V(\tilde h) = \delta\bar V(\bar h)$, $\delta\tilde K(\tilde h) = \delta\bar K(\bar h)$ and $\delta\tilde P(\tilde h) = \delta\bar P(\bar h)$ and rewrite the above equations in terms of these new fields. A prime on a tilde function is then meant to denote derivation with respect to $\tilde h$, while a prime on a barred function means derivation with respect to $\bar h$. Similarly to the flat case, the first equation above is an integro-differential equation, which we would like to bring into a pure differential form. For this purpose, as before, we divide the equation by $\tilde K'=\tilde f\sin(2\bar h)$ and take the derivative of the equation with respect to $\bar h$ while multiplying it by $\tilde f$ to find a differential equation in terms of $\mathcal{K}(\bar h)\equiv d(\tilde f\delta\bar K/\tilde K')/d\bar h$, $\mathcal{V}\equiv \delta\bar V'$ and $\delta\bar P$
\be \label{flowK_sph_2}
\lambda\, \mathcal{K}  =
\frac{d}{d\bar h}\bigg(\tilde f\delta\zeta_K\,\frac{\tilde{K}}{\tilde{K}'}\bigg) -\frac{\tilde f^2}{2}\delta\zeta_J \,=\, \frac{d}{d\bar h}\!\left(\tilde f^2\delta\zeta_K\,\frac{\tan\bar h}{2}\right) -\frac{\tilde f^2}{2}\,\delta\zeta_J,
\ee
The quantities $\delta\zeta_V$, $\delta\zeta_J$, $\delta\zeta_K$ and $\delta(\tilde P\zeta_P)$ are expressed in terms of the fluctuations as follows
{\setlength\arraycolsep{3pt}
\bea 
\tilde f^2\,\tilde V \,\delta\zeta_V &=& \!\!-b\tilde f^2\delta\bar V -3c_d\cot\bar h \,\delta\bar V' -3c_d \,\delta\bar V'',  \\[2mm]
\tilde f^4 c^{\!-\!1}_d\delta\zeta_J \hspace{-1pt} &=& \!\!-2\csc^4\!\bar h\,(3\delta\bar K+\delta\bar P)+2\cos\bar h\csc^3\!\bar h\,(3\delta\bar K'+\delta\bar P') \nn\\
&& \!\!-\csc^2\!\bar h\,(3\delta\bar K''+\delta\bar P'')-12\cot\bar h \,\delta\tilde V', \\[2mm]
\tilde f^4 c^{\!-\!1}_d\delta\zeta_K \hspace{-2pt} &=& \,2\csc^4\!\bar h\,(2\cos(2\bar h)-1)\,\delta\bar K - 2\cos\bar h\csc^3\!\bar h\,\delta\bar K'-\csc^2\!\bar h\,\delta\bar K'' \nn\\
&& +\csc^4\!\bar h\,(\cos(2\bar h)-3)\,\delta\bar P-\cos\bar h\csc^3\!\bar h\,\delta\bar P' -8\cot\bar h\,\delta\bar V'-4\,\delta\bar V'', \\[2mm]
&& \hspace{-1cm} \tilde f^2c^{\!-\!1}_d\delta(\tilde P\zeta_P) \,=\, 4(1+2\csc^2\!\bar h)\,\delta\bar P+\cot\bar h \,\delta\bar P' -\delta\bar P''.
\eea}%
The explicit form of the linearized equations are found by substituting the above functions $\delta\zeta_V$, $\delta\zeta_J$, $\delta\zeta_K$ and $\delta(\tilde P\zeta_P)$ into (\ref{flowV_sph_1},\ref{flowP_sph_1}) and \eqref{flowK_sph_2}
{\setlength\arraycolsep{3pt}
\bea 
0 &=& \delta\bar V'' +3\cot\bar h\,\delta\bar V'+(b+\lambda)\tilde f^2c^{\!-\!1}_d\,\delta\bar V,  \label{eqn_sph_V} \\[3mm]
0 &=& \mathcal{K}'' +(\cot\bar h-2\tan\bar h)\,\mathcal{K}'+\big(6+\tilde f^2 c^{\!-\!1}_d\lambda -4\csc^2\!\bar h-2\sec^2\!\bar h\big)\,\mathcal{K} \nn\\
&& \!\!\!+2\tan\bar h\,\mathcal{V}''+2(\sec^2\!\bar h +2)\,\mathcal{V}'-6\cot\bar h\,\mathcal{V} \nn\\[1mm]
&& \!\!\!+2(\csc^2\!\bar h +1) \csc(2\bar h)\,\delta\bar P' +2(\sec^2\!\bar h-2\csc^4\!\bar h)\,\delta\bar P, \label{eqn_sph_K} \\[3mm]
0 &=& \delta\bar P''-\cot\bar h\,\delta\bar P'+\big((\lambda+d-2)\tilde f^2 c^{\!-\!1}_d-4(1+2\csc^2\!\bar h)\big)\,\delta\bar P. \label{eqn_sph_P}
\eea}%
In order to find the $O(4)$-breaking solutions, we first need to find the nontrivial solutions of \eqref{eqn_sph_P}. These are given by
\be 
\delta\bar P_i = \sin\bar h \,P_{i}^3(\cos\bar h), \hspace{1cm} \lambda^P_i = c_d(i^2+i+4)/\tilde f^2-d+2, \hspace{4mm} i=3,4,\ldots, \label{P_sol_sph}
\ee 
where the functions $P^m_\ell$ are the associated Legendre polynomials. Generically, these sets of solutions do not have common eigenvalues with the solutions to \eqref{eqn_sph_V}
\be
\delta\bar V_i = \sin^{\!-\!1}\!\bar h \,P_{i}^1(\cos\bar h), \hspace{1cm} \lambda^V_i= c_d(i+2)(i-1)/\tilde f^2-b, \hspace{4mm} i=1,2,3\ldots. \label{V_sol_sph}
\ee
This is true in particular at the fixed point where $\tilde f^2=6c_d/(d-2)$, $b=d$. So when we turn on $\delta P_i$, we can no longer have an eigenperturbation in the potential. We are therefore restricted to the two equations \eqref{eqn_sph_K} and \eqref{eqn_sph_P} with $\delta V=0$. In order to find the solution for $\mathcal{K}$ we rewrite these two equations in the compact form
{\setlength\arraycolsep{2pt}
\bea 
\mathfrak{L}_{K\!K}\,\mathcal{K} + \mathfrak{L}_{K\!P}\,\delta\bar P &=& \lambda\;\mathcal{K}, \label{comp_eq_sph_K}\\
\mathfrak{L}_{P\!P}\,\delta\bar P &=& \lambda \,\delta\bar P. \label{comp_eq_sph_P}
\eea}%
There are two set of solutions to the homogeneous equation $\mathfrak{L}_{K\!K} \mathcal{K} = \lambda \,\mathcal{K}$ 
{\setlength\arraycolsep{2pt} 
\bea 
\mathcal{K}^{\mathrm{hom}}_{1,i} = \cot\bar h\,\csc\bar h\, \phantom{}_{2\!}F_1(-i,i+\frac{1}{2},\frac{5}{2},\cos^2\bar h),\hspace{8mm} \lambda^K_{1,i} = 2c_d(2i^2+i-4)/\tilde f^2, \hspace{4mm} i=2,3,4,\ldots,  \label{K_sol_sph_1} \\
\mathcal{K}^{\mathrm{hom}}_{2,i} = \csc^2\!(2\bar h)\, \phantom{}_{2\!}F_1(-i,i-\frac{5}{2},-\frac{1}{2},\cos^2\bar h),\hspace{8mm} \lambda^K_{2,i} = 2c_d(2i^2-5i-1)/\tilde f^2, \hspace{4mm} i=2,3,4,\ldots.  \label{K_sol_sph_2} 
\eea}%
and the two functions $\delta K_{1,i}$ and $\delta K_{2,i}$ are given in terms of these solutions by
\be  \label{integrated_K}
\delta\bar K_{1(2),i}(\bar h) = \sin(2\bar h)\!\int_0^{\bar h} \!\!\! d\sigma \,\mathcal{K}^{\mathrm{hom}}_{1(2),i}(\sigma).
\ee
The integral in $\delta K_{2,i}$ must be taken with the assumption $0<\bar h<\pi/2$ and extrapolated to the whole region $0<\bar h<\pi$ afterwards. The possible constants of integration can be shown to vanish by the properties of the unintegrated equation \eqref{flowK_sph_1}. 
In order to study the $\delta P\neq 0$ case, we pick a solution $\delta P_i, \,\lambda^P_i$ to \eqref{comp_eq_sph_P} given by \eqref{P_sol_sph} and insert it into \eqref{comp_eq_sph_K}
\be  \label{comp_eq_sph_K_i}
\mathfrak{L}_{K\!K}\,\mathcal{K} + \mathfrak{L}_{K\!P}\,\delta\bar P_i = \lambda^{\!P}_i \,\mathcal{K}.
\ee
The general solution to this equation consists of a solution to the homogeneous version $\mathfrak{L}_{K\!K} \mathcal{K} = \lambda^P_i \,\delta K$ plus any specific solution. It can be seen generically, and in particular at the fixed point, that there are no common eigenvalues between the solutions \eqref{P_sol_sph} and (\ref{K_sol_sph_1},\ref{K_sol_sph_2}). So any solution we find to eq.\eqref{comp_eq_sph_K_i} is the unique one. In order to find this solution, we follow the same idea as that used in the flat case and define the operator 
\be 
D \equiv \frac{d}{d\bar h}\,\frac{1}{\sin(2\bar h)}.
\ee
Using the fact that $\mathcal{K} = D \delta \bar K$, eq.\eqref{comp_eq_sph_K_i} can be re-expressed as
\be
\mathfrak{L}_{K\!K} D\, \delta\bar K + \mathfrak{L}_{K\!P}\,\delta\bar P_i = \lambda^{\!P}_i\, D\, \delta\bar K. \label{comp_eq_sph_K_1}
\ee
The following operator identity can be easily verified
\be  
\mathfrak{L}_{K\!K} D -D\, \mathfrak{L}_{P\!P}=3\mathfrak{L}_{K\!P}-\alpha D, \hspace{1cm} \alpha \equiv 6c_d/\tilde f^2 -d+2.
\ee
Using this identity we can rewrite eq.\eqref{comp_eq_sph_K_1} as
\be
D(\mathfrak{L}_{P\!P}-\lambda^{\!P}_i-\alpha)\,\delta\bar K+\mathfrak{L}_{K\!P}\,(3\,\delta\bar K+\delta\bar P_i)=0. \label{comp_eq_sph_K_2}
\ee
Let us first restrict ourselves to perturbations around the fixed point where $\alpha=0$. In this case the above equation suggests choosing $\delta\bar K$ to be proportional to $\delta\bar P_i$ to make the first term vanish, and choosing the proportionality factor to be $-1/3$ to make the second term vanish. So the unique solution is $\delta\bar K_i = -\delta\bar P_i/3$. In summary, at the fixed point, the eigensolutions with $\delta\bar P\neq 0$ are 
\be \label{P_i_sphere}
\boxed{\delta\bar P_i = \sin\bar h \,P_{i}^3(\cos\bar h), \hspace{5mm}\delta\bar K_i = -\frac{1}{3}\,\sin\bar h \,P_{i}^3(\cos\bar h)\hspace{8mm} \lambda_i = \frac{1}{6}\,(d-2)(i^2+i-2), \hspace{3mm} i=3,4,5,\ldots}
\ee
The eigenvalues are all found to be positive. This means that the perturbations are all irrelevant in the IR. So although these perturbations break the $O(5)$ symmetry down to $SU(2)_L\!\times U(1)_Y$, in the IR $O(5)$ symmetry is restored. The $O(4)$-violating deformations start with a quartic term $\delta\bar P_i = \mathcal{O}(\bar h^4)$, so 
they can be written as a Taylor series in $H^\dagger H$ times the operator $\vert H^\dagger\overset{{\hspace{0.5mm}\raisebox{-0.6mm}{\text{\scriptsize $\leftrightarrow$}}}}{\partial}_{\!\mu}\, H\vert^2$.\\[1mm]
It might be worth mentioning that, as pointed out earlier, when restricting to the spherical geometry, the fixed point is UV attractive. The corresponding UV irrelevant eigenfunction around the fixed point is found by an infinitesimal deformation of the sphere radius in $\tilde f^2\sin^2(\tilde h/\tilde f)$. This actually corresponds to the lowest-order deformation in the $O(4)$-preserving eigensolutions (\ref{K_sol_sph_1},\ref{K_sol_sph_2}), namely $\mathcal{K} =\mathcal{K}^{\mathrm{hom}}_{2,2}$ and $ \lambda=\lambda^K_{2,2} = -6c_d/\tilde f_*^2=2-d$ in agreement with the result of \cite{codello_percacci}.  \\[1mm]
Away from the fixed point where $\alpha\neq 0$, we expect $\delta\bar K_i$ to receive corrections proportional to $\alpha$. The eigenvalues are given by \eqref{P_sol_sph} which, when written in terms of $\alpha$ and the eigenvalues in \eqref{P_i_sphere}, are expressed as $\lambda_i+\alpha(i^2+i+4)/6$. In order to construct the eigenfunctions, we find it more convenient to go back to the original equation \eqref{comp_eq_sph_K_i} and expand $\mathfrak{L}_{K\!P}\,\delta\bar P_{i}$ in the basis of the eigenfunctions \eqref{K_sol_sph_1} or \eqref{K_sol_sph_2} depending on whether $i$ is even or odd, and accordingly choose the appropriate ansatz for $\mathcal{K}$. The situation is summarized as follows
\be  \label{even}
\mathfrak{L}_{K\!K}\,\mathcal{K}_{2i} + \mathfrak{L}_{K\!P}\,\delta\bar P_{2i} = \lambda^{\!P}_{2i} \,\mathcal{K}_{2i}, \hspace{5mm} \mathfrak{L}_{K\!P}\,\delta\bar P_{2i}=\sum_{n=2}^i \gamma^n_{2i} \,\mathcal{K}^{\mathrm{hom}}_{1,n} \hspace{6mm}  \mathcal{K}_{2i} = \sum_{n=2}^i \beta^n_{2i} \,\mathcal{K}^{\mathrm{hom}}_{1,n},
\ee
\be  \label{odd}
\mathfrak{L}_{K\!K}\mathcal{K}_{2i\!+\!1}\! +\! \mathfrak{L}_{K\!P}\delta\bar P_{2i\!+\!1} = \lambda^{\!P}_{2i\!+\!1} \mathcal{K}_{2i\!+\!1}, \hspace{4mm} \mathfrak{L}_{K\!P}\delta\bar P_{2i\!+\!1}=\sum_{n=2}^{i+2} \gamma^n_{2i\!+\!1} \mathcal{K}^{\mathrm{hom}}_{2,n} \hspace{5mm} \mathcal{K}_{2i\!+\!1} = \sum_{n=2}^{i+2} \beta^n_{2i\!+\!1} \mathcal{K}^{\mathrm{hom}}_{2,n}.
\ee
We report here the $\gamma^n_i$ coefficients for the first few lowest-order perturbations
\be
\ba{lll}
\tilde f^2c^{\!-\!1}_d\mathfrak{L}_{K\!P}\,\delta\bar P_{3} &=& 96\,\mathcal{K}^{\mathrm{hom}}_{2,2} + 24\,\mathcal{K}^{\mathrm{hom}}_{2,3}, \\ [1.5mm]
\tilde f^2c^{\!-\!1}_d\mathfrak{L}_{K\!P}\,\delta\bar P_{4} &=& 315\,\mathcal{K}^{\mathrm{hom}}_{1,2}, \\[1.5mm] 
\tilde f^2c^{\!-\!1}_d\mathfrak{L}_{K\!P}\,\delta\bar P_{5} &=& 204\,\mathcal{K}^{\mathrm{hom}}_{2,2} -504\,\mathcal{K}^{\mathrm{hom}}_{2,3}-120\,\mathcal{K}^{\mathrm{hom}}_{2,4}, \\[1.5mm]
\tilde f^2c^{\!-\!1}_d\mathfrak{L}_{K\!P}\,\delta\bar P_{6} &=& \hspace{-3mm} -4410\,\mathcal{K}^{\mathrm{hom}}_{1,2} -3937.5\,\mathcal{K}^{\mathrm{hom}}_{1,3}, \hspace{3mm} \cdots 
\ea
\ee
Having computed the coefficients $\gamma^n_{2i}$ and $\gamma^n_{2i+1}$, equations (\ref{even},\ref{odd}) turn into algebraic equations to be solved for $\beta^n_{2i}$ or $\beta^n_{2i+1}$. The solutions are $\beta^n_{2i}=\gamma^n_{2i}/(\lambda^{\!P}_{2i}\!-\!\lambda^{\!K}_{1,n})$ and $\beta^n_{2i+\!1}=\gamma^n_{2i+\!1}/(\lambda^{\!P}_{2i+\!1}\!-\!\lambda^{\!K}_{2,n})$. Finally one has to use the equations \eqref{integrated_K} to find the corresponding $\delta\bar K_{2i}$ and $\delta\bar K_{2i+1}$. \\[1mm]
The eigenvalue expressions \eqref{P_sol_sph} imply that if the sphere radius $\tilde f$ is small enough, $\tilde f^2 < 16c_d/(d-2)$, which includes the fixed radius $\tilde f_*^2 = 6c_d/(d-2)$ as well, then the $O(4)$-violating deformations are IR stable, while for larger values of the radius the first few lowest-order modes turn unstable. The larger the radius, the more the number of eigenperturbations which turn unstable in the IR.

\section{Summary}

We have used a geometric approach to study the one-loop renormalization group evolution of an electroweak invariant four-scalar theory, where the symmetry is nonlinearly realized. Flow equations for three independent functions were found which incorporate the renormalization group running of infinitely many couplings parametrizing the theory. These flow equations were used to study the stability of flat, cylindrical and spherical geometries 
under $O(4)$-violating perturbations and exact analytic expressions for the spectrum and the corresponding eigenperturbations were found. The flat geometry is a fixed point of the flow equations which is, as expected, IR stable against electroweak invariant deformations which break $O(4)$ symmetry. The cylindrical and spherical geometries are preserved under the renormalization group flow with only a running radius, which is attracted to a fixed point in the UV. For small enough values of the radius , including the fixed point value, the two geometries are IR stable under $O(4)$-violating deformations. In other words, if we start with a cylindrical or spherical geometry and slightly deform the geometry in a direction that breaks $O(4)$ symmetry, the deformations will damp down as we move towards the IR, the symmetries will be restored, and the flow will continue with an evolving radius. \\[1mm]
Although the analyses are performed at the one-loop level, the computational approach we have taken is adapted to the use of functional renormalization group methods which might be used to go beyond perturbation theory. This work therefore lays the basis for future investigations regarding more realistic versions with gauge and fermionic degrees of freedom as well as their nonperturbative studies.

\section*{Acknowledgements}

I would like to thank A. Codello, R. Percacci and O. Zanusso for useful comments on the draft and related discussions. I have also benefited from a discussion with A. Wipf.
 
\appendix

\section{Details on the target space geometry}

The Christoffel symbols corresponding to the metric $G_{ij}$ are
{\setlength\arraycolsep{2pt}
\be  \label{Gamma}
\ba{lll}
\Gamma^{0}_{00} &=& \displaystyle \frac{J'}{2J}  \\
\Gamma^{0}_{0\alpha} &=& 0 
\ea \hspace{1cm}
\ba{lll}
\Gamma^{0}_{\alpha\beta} &=& \displaystyle -\frac{K'}{2J}g_{\alpha\beta}  \\
\Gamma^{\alpha}_{00} &=& 0  
\ea \hspace{1cm}
\ba{lll}
\Gamma^{\alpha}_{0\beta} &=& \displaystyle \frac{K'}{2K}\delta^{\alpha}_{\beta}  \\
\Gamma^{\delta}_{\alpha\beta} &=& \displaystyle \left(\Gamma_g\right)^{\delta}_{\alpha\beta},
\ea
\ee}%
where $\Gamma_g$ denotes the Christoffel symbol for the metric $g_{\alpha\beta}$. Also the quantities $\delta \Gamma^k_{ij} = \tilde\Gamma^k_{ij} -\Gamma^k_{ij}$, defined as the difference between the Christoffel symbols for $\tilde G_{ij}$ and $G_{ij}$, are given by
\be \label{deltaGamma}
\delta \Gamma^0_{ij}  =  - \frac{P'}{2J}\, L^3_i L^3_j, \hspace{1cm}
\delta \Gamma^\gamma_{i0} = \frac{K(PK^{-1})'}{2(K+P)}\, L^\gamma_3 L^3_i, \hspace{1cm}
\delta \Gamma^\gamma_{\alpha\beta} = -\frac{P}{K^2}\,\nabla^\gamma(\mathcal{L}^3_\alpha\mathcal{L}^3_\beta),
\ee
where $\nabla_i$ is the covariant derivative compatible with the metric $G_{ij}$ and by definition $L_I^0=L^I_0=0$. Recall that the indices on $L_I^\alpha$ are raised and lowered with $g_{\alpha\beta}$. We have also defined
\be \label{calL}
\mathcal{L}^i_I \equiv L^i_I, \hspace{1cm}  \mathcal{L}^I_i \equiv G_{ij} \mathcal{L}^j_I,
\ee
with $I=1,2,3$ denoting the label of the vector fields. The quantities $L_I^\alpha$, being left invariant vector fields on $SU(2)$, are Killing vectors of $g_{\alpha\beta}=L^I_\alpha L^I_\beta$. At different stages of the computations we have also made use of the fact that $\mathcal{L}^i_I$ are Killing vectors of $G_{ij}$. This can be seen perhaps most easily by direct computation as follows: Using the Christoffel symbols \eqref{Gamma} and the definition \eqref{calL}, one obtains for the $i,j=\alpha,\beta$ components of the tensor $\nabla_{\!i} \mathcal{L}^I_j$   
\be 
\nabla_{\!\alpha} \mathcal{L}^I_\beta = K(\partial_\alpha L_\beta - (\Gamma_g)^\delta_{\alpha\beta}L^I_\delta)=K\nabla^g_{\!\alpha} L^I_\beta,
\ee
where $\nabla^g_\alpha$ is the covariant derivative compatible with $g_{\alpha\beta}$. The antisymmetric property of $\nabla_{\!\alpha} \mathcal{L}^I_\beta$ then follows from that of $\nabla^g_\alpha L^I_\beta$. Also the $0,\alpha$ and $\alpha,0$ components of $\nabla_{\!i} \mathcal{L}^I_j$ become
\be 
\nabla_{\!0} \mathcal{L}^I_\alpha = \frac{K'}{2}L^I_\alpha, \hspace{1cm} \nabla_{\!\alpha} \mathcal{L}^I_0 = -\frac{K'}{2}L^I_\alpha,
\ee
which sum up to zero. Finally, the $\nabla_{\!0} \mathcal{L}^I_0$ component vanishes because $\Gamma^\alpha_{00}=0$. This proves the claim. The expressions \eqref{deltaGamma} are found using the formula
\be  
\delta \Gamma^k_{ij} = \frac{1}{2}\,\tilde{G}^{km}\left(\nabla_{\!i}\delta G_{mj}+\nabla_{\!j}\delta G_{mi}-\nabla_{\!m}\delta G_{ij}\right), \hspace{1cm} \tilde{G}^{ij}= G^{ij}\!-\!\frac{P}{K(K\!+\!P)}L^i_3L^j_3,
\ee
and the Killing property of $\mathcal{L}_I^\alpha$, where $\tilde{G}^{ij}$ is the inverse of $\tilde{G}_{ij}$ and $\delta G_{ij}=\tilde G_{ij}-G_{ij}$. Another useful identity which is used in the computations is 
\be \label{khosro}
\nabla_{\!\alpha}\mathcal{L}^\rho_3\,\nabla_{\!\beta}\mathcal{L}^3_\rho = K\left(g_{\alpha\beta} - L^3_\alpha L^3_\beta\right),
\ee
where, again, in its derivation, the fact that $\mathcal{L}_I^\alpha$ is a Killing vector is used. Recall also that the index $\rho$ runs over $1,2$ and $3$. With the aid of the Riemann tensor for the $O(4)$ invariant metric, reported in \cite{percacci_safari}, and the following formula for the difference between the Riemann tensors of $\tilde G_{ij}$ and $G_{ij}$ 
\be
\delta R_{ij\phantom{k}l}^{\phantom{ij}k} = 2 \nabla_{[i}\delta \Gamma^k_{j]l} +2\delta \Gamma^k_{[i\vert m\vert} \delta \Gamma^m_{j]l}, \hspace{1cm} \tilde R_{ij\phantom{k}l}^{\phantom{ij}k} = R_{ij\phantom{k}l}^{\phantom{ij}k}+\delta R_{ij\phantom{k}l}^{\phantom{ij}k},
\ee
and using the expressions for the Christoffel symbols \eqref{Gamma} and \eqref{deltaGamma}, the identity \eqref{khosro} and the Killing properties of $\mathcal{L}_I^\alpha$, one can obtain, with some patience, the following relations regarding the Riemann tensor of \eqref{metric}
{\setlength\arraycolsep{2pt}
\bea
\tilde R_{0\alpha 0\beta} &=& \left[\frac{K'^2}{4K}-\frac{K''}{2}+\frac{K'J'}{4J}\right]\,g_{\alpha\beta}\!+\!\left[\frac{\left(K\!+\!P\right)'^2}{4(K\!+\!P)}\!-\!\frac{K'^2}{4K}\!-\!\frac{P''}{2}\!+\!\frac{P'J'}{4J}\right] L^3_\alpha L^3_\beta, \label{Riem0a0b} \\[1mm]
\tilde R_{\alpha\gamma\beta\delta}\,L^\gamma_3L^\delta_3 &=& \left[\frac{(K+P)^2}{K}-\frac{(K+P)'K'}{4J}\right]\left(g_{\alpha\beta}-L_\alpha^3 L_\beta^3\right), \label{RiemL3L3} \\[2mm]
\hspace{-8mm}\tilde R_{\alpha\gamma\beta\delta}\,(L^\gamma_1L^\delta_1\!+\!L^\gamma_2L^\delta_2) &=& \left[\frac{4KJ\!-\!K'^2}{4J}\!-\!3P\right] g_{\alpha\beta}\!+\!\left[\frac{4KJ\!-\!K'^2}{4J}\!+\!7P\!+\!\frac{2P^2}{K}\!-\!\frac{K'P'}{2J}\right]L_\alpha^3 L_\beta^3, \label{RiemL12L12}
\eea}%
which are used to find the numerators in \eqref{TrPMPR}. The nonzero components of the Ricci tensor are then easily obtained from the three identities above
{\setlength\arraycolsep{3pt}
\bea
\tilde{R}_{00} &=& \frac{K'^2}{2K^2}-\frac{K''}{K}+\frac{K'J'}{2KJ}+\frac{(K+P)'^2}{4(K+P)^2}-\frac{(K+P)''}{2(K+P)}+\frac{(K+P)'J'}{4(K+P)J}, \label{Ricci00} \\[3mm]
\tilde{R}_{\alpha\beta} &=& \left[2-\frac{2P}{K}-\frac{K''}{2J}+\frac{K'J'}{4J^2}-\frac{K'(K+P)'}{4J(K+P)}\right]g_{\alpha\beta} \label{Ricciab} \nn\\[1mm]
&& \hspace{-3.3mm}+\hspace{-0.3mm}\left[\frac{6P}{K}+\frac{2P^2}{K^2}-\frac{P''}{2J}+\frac{P'J'}{4J^2}-\frac{PK'(K+P)'}{2KJ(K+P)}+\frac{P'(K+P)'}{4J(K+P)}\right]L_\alpha^3 L^3_\beta.
\eea}%

\section{Calculation of beta functionals}

In order to compute the terms in the expansion \eqref{trace}, we use the general formula for the trace of a function $W(\Delta)$ of a Laplace-type operator $\Delta$
\be \label{TrW}
\mathrm{Tr}[W(\Delta)] =
\frac{1}{(4\pi)^{\frac{d}{2}}} \sum_{n= 0}^{\infty} B_{2n}(\Delta)\, Q_{\frac{d}{2}-n}(W).
\ee
The factors $B_{2n}$ are the coefficients which appear in the heat kernel expansion
\be \label{hke}  
\mathrm{Tr}\left(e^{-s\Delta}\right) = \frac{1}{(4\pi)^{\frac{d}{2}}} \sum_{n= 0}^{\infty} B_{2n}(\Delta) s^{-\frac{d}{2}+n},
\ee
and the $Q$-functionals, for non-negative integer $n$, are given by the Mellin transform of $W$
\be 
Q_n(W)=\frac{1}{\Gamma(n)}\int_{0}^{\infty}\!\!\! dz\, z^{n-1} W(z).
\ee
For convenience the optimized cutoff $R_{k}(z)=(k^2-z)\theta(k^2-z)$ of
\cite{litim_optimized} has been used in the computations, which results in the following simple expression for the $Q$-functionals
\be
Q_n \bigg[\frac{\dot{R}_{k}}{(P_k +q)^l}\bigg] =
\frac{2 k^{2(n-l+1)}}{\Gamma(n+1)(1 +\tilde{q})^l}, \hspace{1cm} q=k^2\tilde q.
\ee
where $q$ is an arbitrary function and $\tilde q$ its dimensionless version. What we need is essentially the $n=0$ term in the sum \eqref{TrW}
\be 
\frac{1}{(4\pi)^{d/2}}\, B_{0}(\Delta)\,Q_{\frac{d}{2}}(W) = \frac{1}{(4\pi)^{d/2}}\!\int \! d^dx \, \mathrm{Tr}\,Q_{\frac{d}{2}}(W),
\ee
which we have denoted by $\mathrm{Tr}_0[W(\Delta)]$ in eqs.(\ref{TrPR},\ref{TrPMPR}) below. For more details on trace techniques refer to the appendix of \cite{cpr}. We now have all the ingredients to compute the beta functionals of $V(\rho)$, $J(\rho)$, $K(\rho)$ and $P(\rho)$. The contribution to the flow of the potential comes from the zero-order term in the expansion \eqref{trace} which is (one half) the trace of the operator 
\be 
(\mathcal{P}^{-1}\dot{\mathcal{R}}_k)^i_j  =  \frac{\dot{R}_{k}\;\delta_{0}^{i}\,\delta^{0}_{j}}{P_k+V''/J-V'J'/2J^2} + \frac{\dot{R}_{k}\;(L^i_1L^1_j+L^i_2L^2_j)}{P_k + V'K'/2KJ} +\frac{\dot{R}_{k}\;L^i_3L^3_j}{P_k +V'(K+P)'/2(K+P)J}.
\ee
In fact it is the $B_0$ term in the trace of the above expression which gives the beta functional of the potential
\be \label{TrPR}
\frac{1}{2}\,\mathrm{Tr}_0[\mathcal{P}^{-1}\dot{\mathcal{R}}_k] = c_d k^d\! \int \! d^dx \left[\frac{1}{1+\tilde{V}''}+\frac{2}{1+ \tilde{V}'\tilde{K}'/2\tilde{K}}+\frac{1}{1+\tilde{V}(\tilde{K}+\tilde{P})'/2(\tilde{K}+\tilde{P})}\right],
\ee
where the result is presented in terms of the tilde functions defined in sec.(\ref{flow_eqns}). The first-order term in the expansion \eqref{trace} contributes to the running of $J(\rho)$, $K(\rho)$ and $P(\rho)$. To find it, we need the operator
{\setlength\arraycolsep{3pt}
\bea
(\tilde M\mathcal{P}^{-1}\dot{\mathcal{R}}_k \mathcal{P}^{-1})_i^j &=& \frac{\dot{R}_{k}\;\tilde{M}_{im}\delta_{0}^{m} \delta_{0}^{j}}{J(P_k+V''/J-V'J'/2J^2)^2}+\frac{\dot{R}_{k}\;\tilde{M}_{im}(L^{m}_1 L^{j}_1 +L^{m}_2 L^{j}_2)}{K(P_k + V'K'/2KJ)^2} \nn\\
&+& \frac{\dot{R}_{k}\;\tilde{M}_{im} L^{m}_3 L^{j}_3}{(K+P)(P_k +V'(K+P)'/2(K+P)J)^2}.
\eea}%
Taking (one half the $B_0$ term of) the trace, one obtains
{\setlength\arraycolsep{3pt}
\bea \label{TrPMPR}
\frac{1}{2}\,\mathrm{Tr}_0[\mathcal{P}^{-1} \tilde M \mathcal{P}^{-1}\dot{\mathcal{R}}_k] &=& c_d k^{d-2}\! \int \! d^dx \left[\frac{J^{-\!1}\tilde{M}_{00}}{(1+\tilde{V}'')^2}+\frac{K^{-\!1}\tilde{M}_{ij}(L^i_1L^j_1+L^i_2L^j_2)}{(1+ \tilde{V}'\tilde{K}'/2\tilde{K})^2} \right. \nn\\
&& \left. \hspace{4cm} \;+\,\frac{(K+P)^{-\!1}\tilde{M}_{ij}L^i_3L^j_3}{(1+\tilde{V}'(\tilde{K}+\tilde{P})'/2(\tilde{K}+\tilde{P}))^2}\right].
\eea}%
The three numerators in \eqref{TrPMPR} are found using the expressions (\ref{Riem0a0b}-\ref{RiemL12L12}) for the Riemann tensor
{\setlength\arraycolsep{2pt}
\bea
\tilde{M}_{00} &=& \left[\frac{K'^2}{4K}-\frac{K''}{2}+\frac{K'J'}{4J}\right]\partial^\mu \varphi^\alpha \partial_\mu \varphi^\beta g_{\alpha\beta}  \nn\\
&-& \left[\frac{K'^2}{4K}-\frac{(K+P)'^2}{4(K+P)}+\frac{P''}{2}-\frac{P'J'}{4J}\right]\partial^\mu \varphi^\alpha \partial_\mu \varphi^\beta L^3_\alpha L^3_\beta, \\[4mm]
\tilde{M}_{\alpha\beta} L^\alpha_3 L^\beta_3 &=& \left[\frac{(K+P)^2}{K}-\frac{(K+P)'K'}{4J}\right]\partial_\mu\varphi^\rho \partial^\mu\varphi^\sigma\left(g_{\sigma\rho} - L^3_\sigma L^3_\rho\right) \nn\\
&+& \left[\frac{(K+P)'^2}{4(K+P)}-\frac{(K+P)''}{2}+\frac{(K+P)'J'}{4J}\right]\partial_\mu\varphi^0 \partial^\mu\varphi^0, \\[4mm]
\tilde{M}_{\alpha\beta}(L^\alpha_1 L^\beta_1 +L^\alpha_2 L^\beta_2) &=& \left[\frac{K'^2}{2K}-K''+\frac{K'J'}{2J}\right]\partial_\mu\varphi^0 \partial^\mu\varphi^0+\left[\frac{4KJ-K'^2}{4J}-3P\right]\partial_\mu\varphi^\rho \partial^\mu\varphi^\sigma g_{\sigma\rho} \nn\\
&+& \left[\frac{4KJ-K'^2}{4J}-\frac{K'P'}{2J}+7P+\frac{2P^2}{K}\right] \partial_\mu\varphi^\rho \partial^\mu\varphi^\sigma L_\sigma^3 L^3_\rho. 
\eea}%
Extracting the coefficients of $\partial_\mu\varphi^0\partial^\mu\varphi^0$, $\partial_\mu\varphi^\alpha\partial^\mu\varphi^\beta g_{\alpha\beta}$ and $\partial_\mu\varphi^\alpha\partial^\mu\varphi^\beta L^3_\alpha L^3_\beta$ in \eqref{TrPMPR}, we find the beta functionals of $J(\rho)$, $K(\rho)$ and $P(\rho)$, respectively. The corresponding zeta functionals are reported in sec.(\ref{flow_eqns}).


\begin{thebibliography}{99}

\bibitem{ccwz}
S.~R. Coleman, J.~Wess and B.~Zumino, ``Structure of phenomenological Lagrangians. 1.'', Phys. Rev. 177 (1969) 2239-2247; C.~G. Callan, Jr., S.~R. Coleman, J.~Wess and B.~Zumino, ``Structure of phenomenological Lagrangians. 2.'', Phys. Rev. 177 (1969) 2247-2250.

\bibitem{percacci_safari}
R. Percacci and M. Safari, ``Functional renormalization of $N$ scalars with $O(N)$ invariance'',  Phys.Rev. D88 (2013) 085007, [arXiv:1306.3918 [hep-th]].

\bibitem{mchm}
K. Agashe, R. Contino and A. Pomarol,
``The Minimal composite Higgs model'', 
Nucl.Phys. B719 (2005) 165-187,
[arXiv:hep-ph/0412089].

\bibitem{fptv}
M. Fabbrichesi, R. Percacci, A. Tonero and L. Vecchi,
``The Electroweak $S$ and $T$ parameters from a fixed point condition'',
Phys.Rev.Lett. 107 (2011) 021803, [arXiv:1102.2113 [hep-ph]].

\bibitem{wilson}
K. Wilson and J. Kogut,
``The Renormalization group and the epsilon expansion'',
Phys. Rep. 12 (1974) 75-200; 
K. Wilson, 
``The Renormalization Group: Critical Phenomena and the Kondo Problem'', 
Rev. Mod. Phys. 47 (1975) 773.

\bibitem{wetterich_eqn}
C. Wetterich,
``Exact evolution equation for the effective potential'',
Phys.Lett. B301 (1993) 90-94.

\bibitem{morris_erg}
T. R. Morris, 
``The Exact renormalization group and approximate solutions'', 
Int.J.Mod.Phys. A9 (1994) 2411-2450.

\bibitem{btw}
J. Berges, N. Tetradis and C. Wetterich, 
``Nonperturbative renormalization flow in quantum field theory and statistical physics'',
Phys.Rept. 363 (2002) 223-386.

\bibitem{tetradis_etal}
N. Tetradis and C. Wetterich,
``Critical exponents from the effective average action'',
Nucl.Phys. B422 (1994) 541-592;  
%
J. Berges, N. Tetradis and C. Wetterich,
``Critical equation of state from the average action '', 
Phys.Rev.Lett. 77 (1996) 873-876.

\bibitem{scheme_dependence_frg}
A. Codello, M. Demmel and O. Zanusso,
``Scheme dependence and universality in the functional renormalization group'', Phys.Rev. D90 (2014) 027701,
[arXiv:1310.7625 [hep-th]].

\bibitem{pion_lagrangian}
J.~Honerkamp, F.~Krause and M.~Scheunert, ``On the equivalence of standard and covariant perturbation series in non-polynomial pion lagrangian field theory'', Nucl.Phys. B69 (1974) 618-636.

\bibitem{ab_epsilon}
G. Altarelli and R. Barbieri, 
``Vacuum polarization effects of new physics on electroweak processes'',
Phys.Lett. B253 (1991) 161-167.

\bibitem{honerkamp1972}
J. Honerkamp, 
``Chiral multiloops'',
Nucl.Phys. B36 (1972) 130-140.

\bibitem{afm}
L. Alvarez-Gaume, D. Z. Freedman and S. Mukhi, 
``The Background Field Method and the Ultraviolet Structure of the Supersymmetric Nonlinear Sigma Model'',
Annals Phys. 134 (1981) 85.

\bibitem{boulware_brown}
D. G. Boulware and L. S. Brown,
``Symmetric Space Scalar Field Theory'',
Annals Phys. 138 (1982) 392.

\bibitem{hps}
P. S. Howe, G. Papadopoulos and K.S. Stelle,
``The Background Field Method and the Nonlinear $\sigma$ Model'',
Nucl.Phys. B296 (1988) 26.

\bibitem{codello_percacci}
A. Codello and R. Percacci,
``Fixed Points of Nonlinear Sigma Models in $d>0$ '',
Phys.Lett. B672 (2009) 280-283,
[arXiv:0810.0715 [hep-th]].

\bibitem{percacci_zanusso}
R. Percacci and O. Zanusso,
``One loop beta functions and fixed points in Higher Derivative Sigma Models'',
Phys.Rev. D81 (2010) 065012, [arXiv:0910.0851 [hep-th]].

\bibitem{fptz}
M. Fabbrichesi, R. Percacci, A. Tonero and O. Zanusso,
``Asymptotic safety and the gauged $SU(N)$ nonlinear $\sigma$-model'',
Phys.Rev. D83 (2011) 025016, [arXiv:1010.0912 [hep-ph]].

\bibitem{bfptv}
F. Bazzocchi, M. Fabbrichesi, R. Percacci, A. Tonero and L. Vecchi,
``Fermions and Goldstone bosons in an asymptotically safe model '',
Phys.Lett. B705 (2011) 388-392, [arXiv:1105.1968 [hep-ph]].

\bibitem{fzw}
R. Flore, A. Wipf and O. Zanusso,
``Functional renormalization group of the non-linear sigma model and the $O(N)$ universality class'',
Phys.Rev. D87 (2013) 065019, [arXiv:1207.4499 [hep-th]].

\bibitem{litim_optimized}
D. F. Litim, 
``Critical exponents from optimized renormalization group flows'',
Nucl.Phys. B631 (2002) 128-158,
[arXiv:hep-th/0203006].

\bibitem{friedan_nl}
D. Friedan, ``Nonlinear Models in Two $+$ Epsilon Dimensions'', 
Phys.Rev.Lett. 45 (1980) 1057;
``Nonlinear Models in Two $+$ Epsilon Dimensions'', 
Annals Phys. 163 (1985) 318.

\bibitem{cpr}
A. Codello, R. Percacci and C. Rahmede,
``Investigating the Ultraviolet Properties of Gravity with a Wilsonian Renormalization Group Equation'',
Annals Phys. 324 (2009) 414-469, [arXiv:0805.2909 [hep-th]].


\end{thebibliography}
\end{document}